\documentclass[notoc]{JHEP3} 

\usepackage{amsmath}
\usepackage{textcomp}
\usepackage{epsfig}

\newcommand{\Fig}[1]{Fig.\,#1}
\renewcommand{\not}[1]{#1 \hskip-0.475em /}

\title{Angular distribution of thrust axis with power-suppressed contribution in
$e^{+}e^{-}$ annihilation}

\author{Kaoru Hagiwara$^{1,2}$, Grisha Kirilin$^{1,3}$\\
\textit{\small ${}^{1}$ KEK Theory Center, Tsukuba, 305-0801 Japan}\\
\textit{\small ${}^{2}$ Sokendai, Tsukuba, 305-0801 Japan}\\
\textit{\small ${}^{3}$ Budker Institute of Nuclear Physics, 630090, Novosibirsk,
Russia}}

\received{\today} 		

\preprint{KEK-TH-1372}	

\abstract{Structure function of $e^{+}e^{-}\to hadrons$ cross section
proportional to the longitudinal part of the hadron tensor is power
suppressed with respect to an event shape variable in the two-jet
region. In the SCET framework, we study the event shape distribution
for this structure function to NLL level of accuracy. As, a result we
obtain the angular distribution of hadron jets as a function of the
thrust, in the two jet region. We further examine effects of
non-perturbative hadronization corrections by adopting a shape function
that reproduce the observed event shape distributions. Impacts of our
findings on the electroweak measurements via the jet angular
forward-backward asymmetry are discussed.}


\begin{document}

\section{Introduction}

An intriguing feature of the Standard Model (SM) is the unification of
weak and electromagnetic interactions. One of the most important
predictions of the SM is that the interactions of all electroweak gauge
bosons are determined by the electromagnetic coupling constant $\alpha$
and one additional parameter -- the weak mixing parameter
$\sin^2\theta_{W}$.

Study of the Z boson pole at $e^{+}e^{-}$ colliders provides the most
accurate determination of the electroweak interaction parameters. The
average value of $\sin^{2}\theta_{W}$ found in experiments at LEP is
$0.23153\pm0.00016$ (for a review and bibliography the reader is
referred to \cite{LEP2006}). An experimental method to measure
$\sin^2\theta_{W}$ is based on the fact that weak isospins for
left-handed and right-handed fermions are not the same. This difference
of coupling constants leads to various angular and polarization
asymmetries. The most accurate measurement of $\sin^{2}\theta_{W}$
comes from the forward-backward asymmetry with flavor tagging of the
final-state quark. The plane orthogonal to the colliding beam lines
divides the space of all direction into two hemispheres. The electron
beam is pointing to the \textquotedblleft forward\textquotedblright\
hemisphere. The asymmetry is defined as the relative difference in the
numbers of events with a reference direction in the \textquotedblleft
forward\textquotedblright\ or \textquotedblleft
backward\textquotedblright\ hemispheres. Usually the thrust axis
supplemented with the direction and charge of a prompt lepton from
charm or bottom meson semi-leptonic decay is adopted as the reference
direction.

Extraction of $\sin^2\theta_{W}$ with high precision requires taking
into account a large number of processes accompanying a basic $e^{+}e^{-}%
\rightarrow Z\rightarrow q\bar{q}$ process. One of them is QCD
interaction in the final state. In order to reduce QCD radiation of
additional partons, one has to implement experimental cuts which
effectively select events in the two-jet region \cite{Abbaneo1998}. The
event-shape variable thrust ($T$) can be used to establish such cuts
\cite{Djouadi1995}. Although, from theoretical point of view the thrust
distribution is known to unprecedented level of accuracy
\cite{Becher2008}, up to now the study of the angular distribution of
the thrust axis depending on thrust value has not been done beyond the
first nontrivial order of perturbation theory \cite{Lampe1993}. It was
noted in Ref.\cite{Lampe1993} that an additional contribution to the
event shape, which changes the~angular distribution, is power
suppressed in the $T\rightarrow1$ limit.

The reason why the thrust axis is so stable in the two-jet region is
the following: by definition, the thrust axis coincides with the total
momentum of final state hadrons in a certain hemisphere, thus the
multiple branchings, leaving secondary hadrons in the same hemisphere,
do not change the thrust axis. For the same reason, it is rather hard
to estimate the influence of multiple hadron radiation on the angular
distribution in the strong two-jet limit using the present Monte-Carlo
event generators. In this paper, we consider in detail the angular
distribution of the thrust axis depending on thrust value taken in the
two-jet region.

The paper is organized as follows. In the next section, we consider
possible mechanisms for QCD radiations to have an effect on the angular
distribution in the two-jet region. In Sect.\ref{sec:pert}, using the
method of expanding by regions, we find the perturbative correction to
the structure function, which corresponds to the so-called longitudinal
part of the hadronic tensor. In sections \ref{sec:form} and
\ref{sec:resum}, we use the SCET (Soft Collinear Effective Theory)
framework to study event shape for this structure function, which
appears due to a local three-body operator in the effective theory.
Comparison with the existing experimental data for the angular
distribution and the forward-backward asymmetry is left to the last
section.

\section{Mechanisms to change the angular distribution\label{sec:mech}}

We restrict our attention to the case where $Z$-boson is the only
intermediate state in the processes $e^{+}e^{-}\rightarrow hadrons$.
The thrust ($T$) dependent cross section for the process
$e^{+}e^{-}\rightarrow hadrons$
is a contraction of the leptonic tensor and the hadronic one:%
\begin{equation}
\int^{1}_{1-\tau}\mathrm{d}T\,\frac{\mathrm{d}\sigma }{\mathrm{d}\cos\theta_{T} \mathrm{d}T}%
=\frac{\alpha^{2} \pi N_{\mathrm{c}}}{2}\frac{Q^{2}}{\left(
Q^{2}-M_{Z}^{2}\right)  ^{2}+M_{Z}^{2}\Gamma_{Z}^{2}}\,L^{\mu\nu}\left(
\mathbf{n}_{e}\right)  H_{\mu\nu}\left(  \tau,\mathbf{n}_{T}\right)  ,
\label{eq:SigmaT}%
\end{equation}
where $Q^{2}=\left(  p_{e^{-}}+p_{e^{+}}\right)  ^{2}$,
$\mathbf{n}_{e}=\mathbf{p}_{e^{-}}/\vert\mathbf{p}_{e^{-}}\vert$, and
$\mathbf{n}_{T}$ is the direction of the thrust axis in the hemisphere
which contains a quark. If one neglects
the electroweak radiative corrections and the lepton masses, the leptonic tensor reads%
\begin{equation}
L^{\mu\nu}\left(  \mathbf{n}_{e}\right)  =\left(  g_{al}^{2}+g_{vl}%
^{2}\right)  g_{\perp}^{\mu\nu}\left(  \mathbf{n}_{e}\right)
  -2ig_{al}g_{vl}a^{\mu\nu}\left(  \mathbf{n}_{e}\right)  ,
\label{eq:LepTen}%
\end{equation}
while the hadronic tensor can be parameterized as follows%
\begin{equation}
H^{\mu\nu}=(g_{vq}^{2}+g_{aq}^{2})\left\{  F\left(  \tau\right)  g_{\perp}^{\mu\nu}\left(  \mathbf{n}_{T}\right)  +2G\left(
\tau\right)  g_{\parallel}^{\mu\nu}\left(  \mathbf{n}_{T}\right)  \right\}
+2ig_{vq}g_{aq}K\left(  \tau\right)  a^{\mu\nu}\left(  \mathbf{n}_{T}\right)
, \label{eq:HadTen}%
\end{equation}
where the coupling constants have the following form%
\begin{align}
g_{al}  &  =g_{ad}=-g_{au}=-\frac{1}{2\sin2\theta_{W}},\qquad g_{vl}%
=g_{al}\left(  1-4\sin^{2}\theta_{W}\right)  ,\notag\\
g_{vu}  &  =-g_{al}\left(  1-\frac{8}{3}\sin^{2}\theta_{W}\right)  ,\qquad
g_{vd}=g_{al}\left(  1-\frac{4}{3}\sin^{2}\theta_{W}\right).
\end{align}
The tensors in Eqs.\,(\ref{eq:LepTen}) and (\ref{eq:HadTen}) are
defined as follows:
\begin{subequations}
\begin{align}
g_{\perp}^{\mu\nu}\left(  \mathbf{u}\right)   &  =-g^{\mu\nu}+\frac{n^{\mu
}\left(  \mathbf{u}\right)  n_{+}^{\nu}\left(  \mathbf{u}\right)  +n^{\nu
}\left(  \mathbf{u}\right)  n_{+}^{\mu}\left(  \mathbf{u}\right)  }%
{2},\\
g_{\parallel}^{\mu\nu}\left(  \mathbf{u}\right)   &  =\frac{1}{4}\left[
n^{\mu}\left(  \mathbf{u}\right)  -n_{+}^{\mu}\left(  \mathbf{u}\right)
\right]  \left[  n^{\nu}\left(  \mathbf{u}\right)  -n_{+}^{\nu}\left(
\mathbf{u}\right)  \right]  ,\\
a^{\mu\nu}\left(  \mathbf{u}\right)   &  =\frac{1}{2}\,\epsilon^{\mu\nu
\alpha\beta}n_{\alpha}\left(  \mathbf{u}\right)  n_{\beta+}\left(
\mathbf{u}\right) ,
\end{align}
\end{subequations}
where
\begin{equation}
\mathbf{u}^{2}=1,\qquad n=\left(  1,-\mathbf{u}\right)  ,\qquad n_{+}=\left(
1,\mathbf{u}\right)  . \label{eq:nnp}%
\end{equation}
The expression (\ref{eq:HadTen}) is the most general parametrization of
the tensor which satisfies
\begin{equation}
\left(  n+n_{+}\right)  _{\mu}H^{\mu\nu}=\left(
n+n_{+}\right)  _{\nu}H^{\mu\nu}=0.\label{eq:n0trans}
\end{equation}
Since we neglect the electron mass and since the hadronic tensor must
be contracted with the leptonic one, only those structure functions
that satisfy the condition (\ref{eq:n0trans}) contribute to the cross
section even for massive quarks.

Let the $Z$-boson be polarized along the electron beam direction in the
$e^{+}e^{-}$ collision rest frame, i.e., it is produced in the the
polarized state $\left\vert 1,m\right\rangle$ with $m=\pm 1$. We then
obtain the following angular distribution:
\begin{multline}
\int_{1-\tau}^{1}\frac{\mathrm{d}\sigma}{\mathrm{d}T
\mathrm{d}\cos\theta_{T}}\sim(g_{vq} ^{2}+g_{aq}^{2})\left[  F\left(
\tau\right)  \left(  1+\cos^{2}\theta _{T}\right)  +2G\left(
\tau\right)  \sin^{2}\theta_{T}\right] \\ -m\,2g_{vq} g_{aq}K\left(
\tau\right)  2\cos\theta_{T}, \label{eq:AngDistr}
\end{multline}
where $\cos\theta_{T}=\mathbf{n}_{T}\cdot\mathbf{n}_{e}$.

\FIGURE[b]{
\parbox[c]{.32\textwidth}{\centering
\epsfig{file=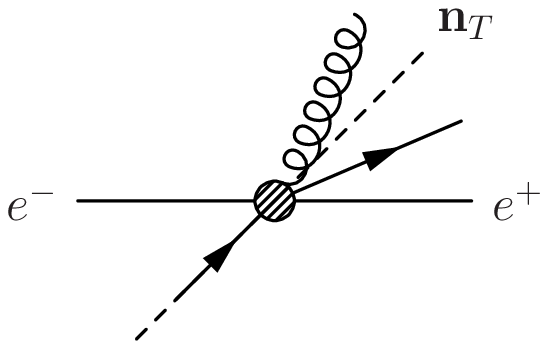,width=.28\textwidth}}
\parbox[c]{.32\textwidth}{\centering
\epsfig{file=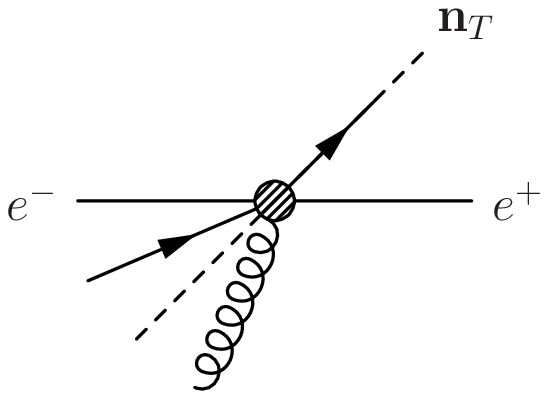,width=.28\textwidth}}
\parbox[c]{.32\textwidth}{\centering
\epsfig{file=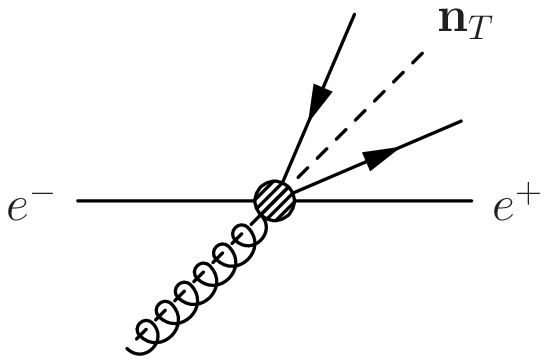,width=.28\textwidth}}\newline(a)
\hspace{.33\textwidth} (b) \hspace{.33\textwidth} (c)
\caption{Three possible directions of the thrust axis in $e^{+}e^{-}\to q\bar{q}g$ processes.}%
\label{fig:topo}}
Let us consider a quark-antiquark ($q\bar{q}$) pair in the final state
in the massless quark limit. In the center-of-mass frame, this pair
should be produced in the spherical helicity state
$d_{\lambda,m}^{1}\left(\mathbf{n}_{q}\right)
=(1+\lambda\,m\cos\theta_{q})/2$, where $\theta_{q}$ is the angle
between $\mathbf{n}_{e}$ and the quark momentum $\mathbf{p}_{q}$, and
$\lambda$=$\pm1$ is the projection of the total spin of the pair on the
direction of $\mathbf{p}_{q}$. Violation of $P$-parity by interaction
with the virtual boson implies different coupling constants for
left-handed and right-handed fermions:
\begin{equation}
g_{vq}-g_{aq}\gamma_{5}=\left(  g_{vq}+g_{aq}\right)  \frac{1}{2}\left(
1-\gamma_{5}\right)  +\left(  g_{vq}-g_{aq}\right)  \frac{1}{2}\left(
1+\gamma_{5}\right)  .
\end{equation}
It results in the following angular distribution for the primary $q\bar{q}%
$-pair:%
\begin{multline}
\frac{\mathrm{d}\sigma}{\mathrm{d}\cos\theta_{q}}\sim\left(
g_{vq}+g_{aq}\right) ^{2}\left\vert d_{m,m}^{1}\right\vert ^{2}+\left(
g_{vq}-g_{aq}\right) ^{2}\left\vert d_{-m,m}^{1}\right\vert
^{2}\\=\frac{1}{2}\left[  \left( g_{vq}^{2}+g_{aq}^{2}\right)  \left(
1+\cos^{2}\theta_{q}\right) -m\left(
2g_{vq}g_{aq}\right)  2\cos\theta_{q}\right]  . \label{eq:Lowest}%
\end{multline}
Assuming $\theta_{q}\approx\theta_{T}$ and comparing
Eq.(\ref{eq:Lowest}) with the expression (\ref{eq:AngDistr}), we find
that the following relations hold in the $\tau\rightarrow0$ limit:
\begin{align}
F\,\left(  \tau\right)   &  =K\left(  \tau\right)  ,\label{eq:FeqK}\\
G\left(  \tau\right)   &  =0. \label{eq:Geq0}%
\end{align}
These relations are the consequence of the free parton approximation
and they have the same nature as the known Callan-Gross relation
\cite{Callan1969,Bjorken1969} or the large recoil symmetry relation for
heavy-to-light form factors \cite{Charles1999}.

Additional radiation of high energy partons results in violation of the
relations (\ref{eq:FeqK}) and (\ref{eq:Geq0}). Let us consider
radiation of a single gluon with energy $E_{g}\sim Q$. There are three
different possibilities for the thrust axis to lie along the momenta of
the final state partons (see \Fig{\ref{fig:topo}}). The topology
(\ref{fig:topo}c), where the thrust axis aligns the gluon momentum, can
in principle be excluded from the analysis. To do this, simultaneous
tagging of both flavored mesons is required. If one tags only one meson
then the contribution of the topology (\ref{fig:topo}c) should also be
taken into account. Due to CP-invariance, this topology does not
contribute to $K\left( \tau\right)$ but gives the main contribution to
$F\,\left( \tau\right)$ and thereby violates the relation
(\ref{eq:FeqK}). In contrast, the topologies
(\ref{fig:topo}a,\thinspace b) contributes mainly to $G(\tau)$ and
hence lead to the violation of the relation (\ref{eq:Geq0}). In the
present paper, we mainly consider the topologies
(\ref{fig:topo}a,\thinspace b) such that $\left\vert
\theta_{q}-\theta_{\bar{q}}\right\vert \approx\pi$ in the
$\tau\rightarrow0$ limit. We will give a few comments about the
topology (\ref{fig:topo}c) in Sect.\,\ref{sec:form}.

\FIGURE[b]{
\epsfig{file=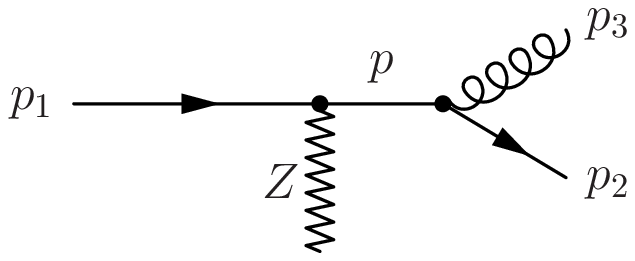,width=.3\textwidth}\hspace{.2\textwidth
}\epsfig{file=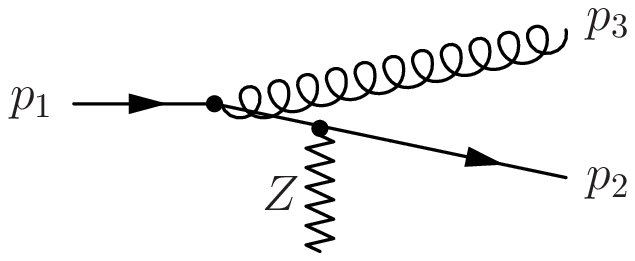,width=.3\textwidth}\\
(a)\hspace{.4\textwidth} (b) \caption{Feynman diagrams for the single
gluon radiation process in hadronic $Z$-boson decays.}
\label{fig:qalong}}

In the covariant perturbation theory, the amplitudes are obtained as
the sum of the two Feynman diagrams, Figs.\,\ref{fig:qalong}(a) and
\ref{fig:qalong}(b). For the sake of completeness, we present here the
leading perturbative results for $F\left( \tau\right) -K\left(
\tau\right)  $ and $G\left(  \tau\right)$ \cite{Lampe1993}:
\begin{multline}
F_{\mathrm{tree}}\left(  \tau\right)  -K_{\mathrm{tree}}\left(
\tau\right) =\frac{\alpha_{s}}{4\pi}C_{\mathrm{F}}\left\{
-\frac{2\pi^{2}}{3}+\frac
{\tau\left(  12\tau^{2}+17\tau-45\right)  }{\tau-1}\right.\\
+\left(  \frac{5}{2}%
-8\ln2-4\tau-2\tau^{2}\right)  \ln(1-2\tau)\\
\left.  +2\tau(\tau+2)\ln\tau+8\ln\left(  1-\tau\right)  \left[ \ln\tau
-\ln(1-2\tau)+6\right]  +8\left[  \operatorname{Li}{}_{2}(\tau
)-\operatorname{Li}{}_{2}(2\tau-1)\right]  \right\}  , \label{eq:FK1}%
\end{multline}
\begin{equation}
G_{\mathrm{tree}}\left(  \tau\right)  =\frac{\alpha_{s}}{\pi}C_{\mathrm{F}%
}\left\{  \tau-4\left[  \frac{\tau\left(  2-\tau\right)  }{1-\tau}+2\ln\left(
1-\tau\right)  \right]  \right\}  , \label{eq:G1}%
\end{equation}
so that the power expansion near $\tau=0$ takes the form:
\begin{align}
F_{\mathrm{tree}}\left(  \tau\right)  -K_{\mathrm{tree}}\left(
\tau\right) & =\frac{\alpha_{s}}{\pi}C_{\mathrm{F}}\left[
\tau\ln\frac{1}{\tau}+O\left( \tau^{2}\right)  \right]  ,\label{eq:SmallTauExpFK}\\
G_{\mathrm{tree}}\left(  \tau\right) & =G^{(0)}\left(  \tau\right)  +O\left(  \tau^{2}\right)  =\frac{\alpha_{s}}%
{\pi}C_{\mathrm{F}}\left[  \tau+O\left(  \tau^{2}\right)  \right]  .
\label{eq:SmallTauExpG}%
\end{align}

\DOUBLEFIGURE[b]{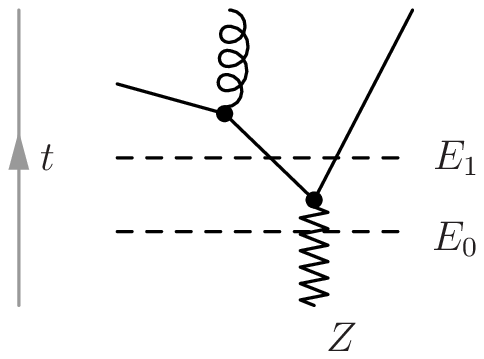,width=.25\textwidth}{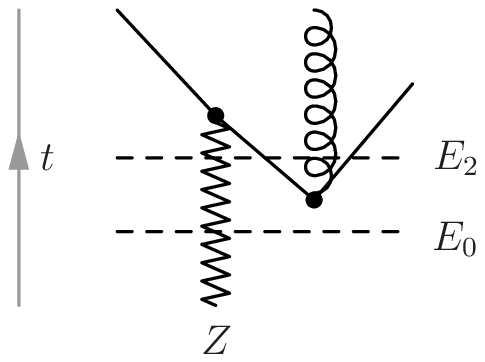,width=.25\textwidth}{Two
particle intermediate state in $oPT$.\label{fig:inter1}}{Four particle
intermediate state in $oPT$.\label{fig:inter2}}

Below we give a qualitative explanation of the physical mechanisms
which results in violation of the relations (\ref{eq:FeqK}) and
(\ref{eq:Geq0}). Let us consider the the topology (\ref{fig:topo}a),
where the gluon is emitted along the quark momentum direction. In order
to remind us of the kinematical configuration concerned, we draw the
diagrams in Figs.\,(\ref{fig:qalong}a) and (\ref{fig:qalong}b) in such
a way that the $q$, $\bar{q}$, $g$ lines follow their momentum
directions.

We find the old-fashioned perturbative theory ($oPT$) useful to
identify two mechanisms that radiation affects the angular
distribution. Using $oPT$, one can decompose each covariant diagram
(\Fig{\ref{fig:qalong}(a)} or \Fig{\ref{fig:qalong}(b)}) into the sum
of two time-ordered diagrams shown in \Fig{\ref{fig:inter1}} and
\Fig{\ref{fig:inter2}}.

The first mechanism corresponds to the time ordering when additional
partons are radiated off the primary $q\bar{q}$-pair \textit{after} the
$Z$-boson decay, as depicted in \Fig{\ref{fig:inter1}}. The
intermediate state involved is real but with the different energy
$E_{1}$, and hance the angular distribution (\ref{eq:Lowest}) is valid
for the relative momentum of the primary $q\bar{q}$-pair. The only
possibility for the subsequent radiation to change the thrust axis
distribution is the following: after the splitting of a parton, one of
the ``child''-parton is radiated into the ``wrong'' thrust hemisphere,
which is opposite to that of its ``parent''-parton. For example, the
$oPT$-decomposition of the diagram (\ref{fig:qalong}b) has such
intermediate and final states. The main effect of such radiation is
that the direction of the relative momentum of the real primary
$q\bar{q}$-pair can differ from the thrust axis.

The second mechanism is when virtual states appears \textit{before} the
$Z$-boson decay, picking up the $Z$-boson from the state
$d_{0,m}^{1}\left(\mathbf{n}_{T}\right)$, as depicted in
\Fig{\ref{fig:inter2}}.

Let us estimate the contribution of the first mechanism, assuming that
$\theta_{q}-\theta_{T}$ misfit is of order $\tau^{1/2}$. For the sake
of brevity, we assume that the $Z$-boson is produced in the state
$\vert1,-1\rangle$. As one can see in \Fig{\ref{fig:qalong}(b)}, the
primary $q\bar{q}$-pair is produced in states
$d_{\pm1,-1}^{1}\left(\mathbf{n}_{q}\right) $, where
$\textbf{n}_{q}=\mathbf{p}_{2}/\vert \mathbf{p}_{2}\vert$, while the
thrust axis $\textbf{n}_{T}$ is along the momentum of the antiquark
$\mathbf{p}_{1}$, i.e. $\textbf{n}_{T}=-\mathbf{p}_{1}/\vert
\mathbf{p}_{1}\vert$. To derive the distribution with respect to
$\theta_{T}$, it is necessary to express $d_{\pm1,-1}^{1}\left(
\mathbf{n}_{q}\right)  $ through $d_{\lambda,-1}^{1}\left(
\mathbf{n}_{T}\right)  $, that leads to the admixture of
$d_{0,-1}^{1}\left(\mathbf{n}_{T}\right) $ state and therefore $G\left(
\tau\right)  \neq0$ as well as $F\,\left( \tau\right) \neq K\left(
\tau\right) $ follow:
\begin{multline}
\mathrm{d}\sigma\sim\left(  g_{vq}+g_{aq}\right)  ^{2}\left\vert d_{-1,-1}%
^{1}\left(  \mathbf{n}_{q}\right)  \right\vert ^{2}+\left(
g_{vq}-g_{aq}\right) ^{2}\left\vert d_{1,-1}^{1}\left(
\mathbf{n}_{q}\right)  \right\vert ^{2}\\%
=\sum_{n,l=-1}^{1}T_{nl}\,d_{n,-1}^{1}\left(\mathbf{n}_{T}\right)
d_{l,-1}^{1}\left(\mathbf{n}_{T}\right)  , \label{eq:SigmaMisfit}%
\end{multline}
with%
\begin{equation}
T_{nl}=\left(  g_{vq}+g_{aq}\right)  ^{2}D_{-1,n}^{\star1}
D_{-1,l}^{1}  +\left(  g_{vq}-g_{aq}\right)  ^{2}%
D_{1,n}^{\star1}  D_{1,l}^{1},
\end{equation}
where $D_{\lambda,m}^{J}$ is the operator of finite rotations:
\begin{equation}
D_{\lambda,m}^{1}\equiv D_{\lambda,m}^{1}(\alpha,\theta_{qT},\gamma).
\end{equation}
Here, $\theta_{qT}$ is the angle between the vectors $\mathbf{n} _{T}$
and $\mathbf{n} _{q}$, $\alpha$ is the angle of rotation about
$\mathbf{n} _{T}$-axis transferring the vector
$\mathbf{n}_{e}\times\mathbf{n} _{T}$ into $\mathbf{n}
_{T}\times\mathbf{n} _{q}$, $\gamma$ is the angle of rotation about
$\mathbf{n} _{q}$-axis transferring the vector $\mathbf{n}
_{T}\times\mathbf{n} _{q}$ into $\mathbf{n}_{e}\times \mathbf{n} _{q}$.
All rotations are defined with respect to the \textquotedblleft right
hand grip rule\textquotedblright. The matrix $T_{nl}$ entering
Eq.(\ref{eq:SigmaMisfit}) does not depend on $\gamma$ and becomes
diagonal after averaging over directions transverse to the thrust axis:
\begin{multline}
\int_{0}^{2\pi}\frac{d\alpha}{2\pi}\,T_{nl}=\left(g_{aq}^{2}+g_{vq}
^{2}\right)\\
 \times\,\mathrm{diag}\left(  \frac{1+\cos^{2}\theta_{qT}}{2}%
+\frac{2g_{aq}g_{vq}}{g_{aq}^{2}+g_{vq}^{2}}\cos\theta_{qT},\sin
^{2}\theta_{qT},\frac{1+\cos^{2}\theta_{qT}}{2}-\frac
{2g_{aq}g_{vq}}{g_{aq}^{2}+g_{vq}^{2}}\cos\theta_{qT}\right)  ,
\end{multline}
and the angular distribution of the thrust axis becomes
\begin{multline}
\mathrm{d}\sigma\sim\frac{1}{2}\left[  \frac{1+\cos^{2}\theta_{qT}}%
{2}\left(  g_{aq}^{2}+g_{vq}^{2}\right)  \left(
1+\cos^{2}\theta_{T}\right)  +\cos\theta_{qT}\left(
2g_{aq}g_{vq}\right)  \left(
2\cos\theta_{T}\right)  \right] \\
+\sin^{2}\theta_{qT}\frac{1}%
{2}\left(  g_{aq}^{2}+g_{vq}^{2}\right)  \sin^{2}\theta_{T}.
\label{eq:1way}%
\end{multline}
The coefficients before the distributions $1+\cos^{2}\theta_{T}$ and
$2\cos\theta_{T}$ are now different, but this difference is strongly
suppressed in the collinear limit $\theta_{qT}\rightarrow 0$:
\begin{equation}
\left.\frac{1+\cos^{2}\theta_{qT}}{2\cos\theta_{qT}}\right\vert
_{\theta_{qT}\rightarrow 0}=1+\frac{\theta_{qT}^{4}}{8}+O\left(\theta_{qT}^{6}\right).
\label{eq:ThetaExp}
\end{equation}
Therefore, the corrections (\ref{eq:1way}) integrated over the region
$\theta_{qT}^{2}\sim\tau$ generate the following
corrections\footnote{Here, we do not distinguish between $\tau^{n}$ and
$\tau^{n}\ln^{m}\tau$, where the logarithmic part can appear due to
integration over the gluon energy.} to the relations (\ref{eq:FeqK}),
(\ref{eq:Geq0}):
\begin{equation}
F\,\left(  \tau\right) -K\left(\tau\right) \sim\tau^{3},\qquad G\sim\tau^{2}. \label{eq:1estim}
\end{equation}

To estimate the effect of virtual states depicted in
\Fig{\ref{fig:inter2}}, let us consider the following projectors
\begin{equation}
\hat{P}_{\left(  +\right)  }\left(  \mathbf{u}\right)  =\frac{\not
n\left(  \mathbf{u}\right)  \not   n_{+}\left(  \mathbf{u}\right)  }%
{4},\qquad\hat{P}_{\left(  -\right)  }\left(  \mathbf{u}\right)
=\frac{\not   n_{+}\left(  \mathbf{u}\right)  \not   n\left(
\mathbf{u}\right)  }{4}, \label{eq:Pdef}%
\end{equation}
where 4-vectors $n$ and $n_{+}$ are defined in Eq.(\ref{eq:nnp}). They
satisfy the following projective relations:%
\begin{equation}
\hat{P}_{\left(  \pm\right)  }^{2}=\hat{P}_{\left(  \pm\right)  },\qquad
\hat{P}_{\left(  \pm\right)  }\hat{P}_{\left(  \mp\right)  }=0,\qquad\hat
{P}_{\left(  +\right)  }+\hat{P}_{\left(  -\right)  }=1.
\end{equation}
To ascertain their geometrical meaning, we consider the following
light-like 4-momenta
\begin{equation}
q=E\left(  1,\,\mathbf{v}\right)  ,\qquad q^{\prime}=E\left(  1,\,\mathbf{n}%
\right)  ,
\end{equation}
and a spinor $u_{\mathbf{q}}^{\left(  \lambda\right)  }$ such that
$\not  qu_{\mathbf{q}}^{\left(  \lambda\right)  }=0$ for a definite
helicity $\lambda$. One can show that the wave functions
$\hat{P}_{\left( \pm\right) }\left(  \mathbf{n}\right)
u_{\mathbf{q}}^{\left(  \lambda \right)  }$ correspond to the two
states with the projection of the angular
momentum $\pm\lambda$ on the axis $\mathbf{n}$, but with the same helicity:%
\begin{equation}
u_{\mathbf{q}}^{\left(  \lambda\right)  }=\hat{P}_{\left(  +\right)
}\,\left(  \mathbf{n}\right)  u_{\mathbf{q}}^{\left(  \lambda\right)  }%
+\hat{P}_{\left(  -\right)  }\,\left(  \mathbf{n}\right)  u_{\mathbf{q}%
}^{\left(  \lambda\right)  }=d_{\lambda,\lambda}^{1/2}\left(  \pi-\theta_{\mathbf{nv}}\right)
u_{-\mathbf{q}^{\prime}}^{\left(  \lambda\right)  }+d_{\lambda,\lambda}^{1/2}\left(  \theta
_{\mathbf{nv}}\right)  u_{\mathbf{q}^{\prime}}^{\left(  \lambda\right)
}, \label{eq:WFExp}%
\end{equation}
where $\cos\theta_{\mathbf{nv}}=\mathbf{n}\cdot\mathbf{v}$. Now let us
consider the fermion propagator inside the Feynman diagram
(\ref{fig:qalong}a). The total momentum of the quark and the gluon is%
\begin{equation}
p^{\mu}=p_{2}^{\mu}+p_{3}^{\mu}=\left(  p\cdot n\right)  \frac{n_{+}^{\mu}}%
{2}+\left(  p\cdot n_{+}\right)  \frac{n^{\mu}}{2}.
\end{equation}
Here and below we omit the arguments of $n^{\mu}$ and $n_{+}^{\mu}$
implying $n^{\mu}\left(  \mathbf{n}_{T}\right)  $ and
$n_{+}^{\mu}\left( \mathbf{n}_{T}\right)  $. Using the projectors
(\ref{eq:Pdef}), one can split
the propagator into two parts%
\begin{equation}
\frac{\not   p}{p^{2}+i0}=\hat{P}_{\left(  -\right)  }%
\frac{\not   p}{p^{2}+i0}\hat{P}_{\left(  +\right)  }+\hat{P}_{\left(  +\right)  }\,\frac{\not
p}{p^{2}+i0}\hat{P}_{\left(  -\right)  }, \label{eq:FeynExp}%
\end{equation}
which correspond to retarded and advanced propagations:
\begin{subequations}
\begin{align}
\hat{P}_{\left(  -\right)  }\,\frac{\not   p}{p^{2}+i0}\hat{P}_{\left(
+\right)  }  &  =\frac{p\cdot n}{p^{2}+i0}\frac{\not  n_{+}}{2}=\frac{1}{E_{0}-E_{1}+i0}\frac{\not   n_{+}}{2}%
,\label{eq:Retarded}\\
\hat{P}_{\left(  +\right)  }\,\frac{\not   p}{p^{2}+i0}\hat{P}_{\left(
-\right)  }  &  =\frac{p\cdot n_{+}}{p^{2}+i0}\frac{\not  n}{2}=\frac{(-1)}{E_{0}-E_{2}-i0}\frac{\not   n}{2},
\label{eq:Advanced}%
\end{align}
\end{subequations}
where the following simple relations were used:
\begin{align}
p^{2} & =(p\cdot n_{+})(p\cdot n),\\
p\cdot n_{+}  &  =p^{0}-\left\vert \mathbf{p}\right\vert =Q-\left(  \left\vert
\mathbf{p}_{1}\right\vert +\left\vert \mathbf{p}\right\vert \right)
=E_{0}-E_{1},\notag\\
-p\cdot n  &  =-p^{0}-\left\vert \mathbf{p}\right\vert =Q-\left(  Q+\left\vert
\mathbf{p}\right\vert +\left\vert \mathbf{p}_{2}\right\vert +\left\vert
\mathbf{p}_{3}\right\vert \right)  =E_{0}-E_{2}.
\end{align}
Here $E_{0}=Q$ is the energy of the initial state, $E_{1}=\left\vert
\mathbf{p}_{1}\right\vert +\left\vert \mathbf{p}\right\vert $ and
$E_{2}=Q+\left\vert \mathbf{p}\right\vert +\left\vert \mathbf{p}%
_{2}\right\vert +\left\vert \mathbf{p}_{3}\right\vert $ are the
energies of the intermediate states depicted in Figs.(\ref{fig:inter1})
and (\ref{fig:inter2}) respectively.
Taking into account the decomposition (\ref{eq:WFExp}), one can see
that the term (\ref{eq:Retarded}) corresponds to the quark propagating
in the direction $\mathbf{n}_{T}$ while the term (\ref{eq:Advanced})
corresponds to the antiquark propagating in the
direction~$-\mathbf{n}_{T}$. Therefore, we can conclude that in the
first diagram (Fig.\ref{fig:inter1}) the $Z$-boson decays into the
quark-antiquark pair with opposite helicities, i.e., from the
$d_{\pm1,-1}^{1}$ state. In the second diagram (Fig.\ref{fig:inter2}),
the $Z$-boson disappears being absorbed by the intermediate antiquark.
Taking into account helicity conservation for massless quarks, we
conclude that it is possible only from the state $d_{0,-1}^{1}$.
According to the energy conservation for the final state, i.e., $\vert
\mathbf{p}_{1}\vert + \vert \mathbf{p}_{2}\vert + \vert
\mathbf{p}_{3}\vert=Q$, we find that $E_{0}-E_{2}=-Q$, i.\,e., the
propagator (\ref{eq:Advanced}) does not depend on the kinematic
configuration of the final state. Since the phase space for a definite
thrust value is of order $\tau$, the intermediate state shown in
\Fig{\ref{fig:inter2}} leads to a contribution
\begin{equation}\label{eq:estim}
G\left( \tau\right) \sim\tau.
\end{equation}
This explains the tree-level result (\ref{eq:SmallTauExpG})).

The analysis presented above remains valid in the case of a
multiparticle final state with primary $q$ and $\bar{q}$ radiated into
the opposite hemispheres. A QCD cascade starting from the
quark-antiquark state as depicted in \Fig{\ref{fig:inter1}} cannot
change the relations (\ref{eq:FeqK}) and (\ref{eq:Geq0}) significantly
in the two jet region, where the corrections are suppressed by
$\tau^{3}$ and $\tau^{2}$ respectively. In the small $\tau$ region, the
leading corrections to the relation (\ref{eq:Geq0}) come from the
process at short distances shown in \Fig{\ref{fig:inter2}}, which gives
rise to the contribution of order $\tau$ (\ref{eq:estim}) . The proper
consideration of violation of the relations (\ref{eq:FeqK}) and
(\ref{eq:Geq0}) requires a consideration of a new type of jets. As it
will be demonstrated below, those jets are initiated by two collinear
partons produced at short distances, i.\thinspace e., due to a local
operator with more than two fields, so that all secondary collinear or
soft particles are radiated coherently by these partons.

\section{Perturbative corrections\label{sec:pert}}

If one has an integral over a scaleless domain whose integrand depends
on a small external parameter $\lambda\ll1$, then \textit{the method of
expanding by regions} gives an asymptotic expansion with respect to
$\lambda$. The series may contain arbitrary non-integer powers of
$\lambda$ as well as integer powers of $\ln\lambda$. Here, we outline
the prescription used below without discussing details of the method,
for which the reader is referred to the original studies
\cite{Smirnov:1998vk, Beneke:1997zp, Smirnov2002}.

The method utilizes the fact that the expansion of an integrand with
respect to $\lambda$ may be invalid in a certain region where the
integrand becomes singular. Analyzing the integrand singularities, one
can establish the so-called \textit{power counting rules}, according to
which one can pick up a simplified singular behavior of the integrand
by expanding it not only in the external parameters but also in
integration variables. Using this method, one can represent the
integral as a sum of integrals, such that each integrand is an
expansion of the original one with respect to the power counting rules.

Let us demonstrate how to apply this method to calculate the leading
perturbative correction to the structure function $F\left( \tau\right)
$ in a region where $1-T<\tau\ll1$:\nopagebreak
\begin{multline}
F\left(  \tau\right)    =\frac{4\pi}{(\mathcal{D}-2)N_{c}Q^{2}}%
\int\mathrm{d}\rho_{_X}
\Theta\left(  \sum_{h\in X}\left\vert \mathbf{p}_{h}\cdot\mathbf{n}%
_{T}\right\vert -(1-\tau)Q\right) \\
\times\sum_{\sigma,c}g_{\perp}^{\mu\nu}\left(  \mathbf{n}%
_{T}\right)  \left\langle X\left\vert \hat{J}_{\nu}\right\vert
0\right\rangle ^{\star}\left\langle X\left\vert
\hat{J}_{\mu}\right\vert 0\right\rangle,
\end{multline}
\begin{equation}
F\left(  \tau\right) =1+\frac{\alpha_{s}}{4\pi}F_{\left(  1\right)  }\left(  \tau\right)
+\left(  \frac{\alpha_{s}}{4\pi}\right)  ^{2}F_{\left(  2\right)  }\left(
\tau\right)  +\ldots.
\end{equation}
where
$\hat{J}_{\mu}=\widehat{\bar{\psi}}_{q}\gamma_{\mu}\widehat{\psi}_{q}$
and $\mathrm{d}\rho_{_X}$ is the phase space of a final state
$\left\vert X\right\rangle $:
\begin{equation}
\mathrm{d}\rho_{_X}=\left(  2\pi\right)  ^{\mathcal{D}}\delta^{\mathcal{D}%
}\left(  Q\,\frac{n+n_{+}}{2}-\sum_{h\in X}p_{h}\right)
{\displaystyle\prod\limits_{h\in X}}
\frac{\mathrm{d}^{\mathcal{D}}p_{h}}{\left(  2\pi\right)  ^{\mathcal{D}-1}%
}\,\delta\left(  p_{h}^{2}\right)  \Theta\left(  p_{h}\cdot n+p_{h}\cdot
n_{+}\right)  \label{eq:PhSp}%
\end{equation}
and $\sum_{\sigma,c}$ denotes the sum over spin and color states and
$\mathcal{D}=4-2\epsilon$ is the space-time dimensionality.

First of all, we introduce a small parameter $\lambda$ such that
$\lambda ^{2}\sim\tau$. We use the Sudakov decomposition to represent
all real or
virtual particle momenta:%
\begin{equation}
p_{h}=\left(  p_{h}\cdot n\right)  \frac{n_{+}}{2}+\left(  p_{h}\cdot
n_{+}\right)  \frac{n}{2}+p_{h\perp}.
\end{equation}
Power counting rules should estimate the components $\left(  p_{h}\cdot
n,p_{h\perp},p_{h}\cdot n_{+}\right)  $ in comparison with $\lambda$.
The regions of integrations and the corresponding power counting rules
are presented in Table \ref{tab:counting}. There are two regions,
namely, hard and soft ones, where all momentum components are of the
same order. In the hard region one should expand an integrand with
respect to $\tau$ only and integrate over real or virtual particle
momenta in dimensional regularization, ignoring any soft or collinear
infrared singularities. The expansion in the soft region corresponds to
an integrand approximation near the soft singularities. There are also
$r$- and $l$-collinear regions, where symmetry among the space
directions is strongly broken. These regions account for the collinear
singularities.
\begin{table}[t]
\centering
\begin{tabular}
[c]{lcc}\hline\hline
Region & Scale & Power counting $Q^{-1}\left(  p_{h}\cdot n,p_{h\perp}%
,p_{h}\cdot n_{+}\right)  $\\\hline
hard & $Q^{2}$ & $\left(  1,1,1\right)  $\\
$r$-collinear & $\tau Q^{2}$ & $\left(  1,\lambda,\lambda^{2}\right)  $\\
$l$-collinear & $\tau Q^{2}$ & $\left(  \lambda^{2},\lambda,1\right)  $\\
soft & $\tau^2 Q^{2}$ & $\left(  \lambda^{2},\lambda
^{2},\lambda^{2}\right)  $\\\hline\hline
\end{tabular}
\caption{Power counting rules}%
\label{tab:counting}%
\end{table}

The contribution of each region is gauge invariant by itself. In
Feynman gauge, virtual corrections to the amplitude $\langle
q\bar{q}\vert \hat{J}_{\mu }\vert 0\rangle $ contribute in the hard
region only, and the
corresponding contribution to $F\left(  \tau\right)  $ has the form:%
\begin{equation}
F_{\left(  1\right)  }^{\mathrm{hard}}\left(  \tau\right)  =C_{\mathrm{F}%
}\left(  \frac{Q^{2}}{\mu^{2}}\right)  ^{-\epsilon}\left(  -\frac{4}%
{\epsilon^{2}}-\frac{6}{\epsilon}-16+\frac{7\pi^{2}}{3}+O\left(
\epsilon\right)  \right)  .\label{eq:R2Hard}%
\end{equation}
In contrast to the virtual corrections, the real emissions give
contributions in the collinear and soft regions:
\begin{equation}
F_{\left(  1\right)  }^{l-\operatorname{col}}\left(  \tau\right)  =F_{\left(
1\right)  }^{r-\operatorname{col}}\left(  \tau\right)  =C_{\mathrm{F}}\left(
\frac{\tau Q^{2}}{\mu^{2}}\right)  ^{-\epsilon}\left(  \frac{4}{\epsilon^{2}%
}+\frac{3}{\epsilon}-\pi^{2}+7+O\left(  \epsilon\right)  \right)+O(\lambda^{2})
,\label{eq:R2Col}%
\end{equation}%
\begin{equation}
F_{\left(  1\right)  }^{\mathrm{soft}}\left(  \tau\right)  =C_{\mathrm{F}}\left(  \frac{\tau^{2}Q^{2}}{\mu^{2}}\right)
^{-\epsilon}\left(  -\frac{4}{\epsilon^{2}}+\frac{\pi^{2}}{3}+O\left(
\epsilon\right)  \right)+O(\lambda^{2})  .\label{eq:R2Soft}%
\end{equation}
The sum of all contributions is a well known result \cite{DeRujula1978}:%
\begin{equation}
F\left(  \tau\right)  =1+\frac{\alpha_{s}}{4\pi}C_{\mathrm{F}}\left(
-4\ln^{2}\frac{1}{\tau}+6\ln\frac{1}{\tau}-2+\frac{2\pi^{2}}{3}\right)
+O\left(  \alpha_{s}\tau\right)  +O\left(  \alpha_{s}^{2}\right)
.\label{eq:R2pert}%
\end{equation}
The singularities with respect to $\epsilon$ and the
$\mu^{2}$-dependence drop out of the sum of all the contributions
(\ref{eq:R2Hard}), (\ref{eq:R2Col}) and (\ref{eq:R2Soft}), which tests
our use of the method of expanding by regions.

Let us apply the same method to calculate the perturbative result for
$G\left(  \tau\right)  $ function in the region $1-T<\tau\ll1$:%
\begin{equation}
G\left(  \tau\right)  =\frac{2\pi}{N_{c}Q^{2}}\int\mathrm{d}\rho_{_X}%
\sum_{\sigma,c}\left\langle X\left\vert \hat{J}_{\parallel}\right\vert
0\right\rangle ^{\star}\left\langle X\left\vert \hat{J}_{\parallel}\right\vert
0\right\rangle \Theta\left(  \sum_{h\in X}\left\vert \mathbf{p}_{h}%
\cdot\mathbf{n}_{T}\right\vert -(1-\tau)Q\right)  \label{eq:Gdef},
\end{equation}
where%
\begin{equation}
\hat{J}_{\parallel}=\widehat{\bar{\psi}}_{q}\left(  \frac{\not n}{2}%
-\frac{\not n_{+}}{2}\right)  \widehat{\psi}_{q}.\label{eq:Jparallel}
\end{equation}

First, we consider the case when the thrust axis is aligned with the
antiquark momentum $p_{1}=E_{1}n$.\ The diagrams we would like to
consider are depicted in \Fig{\ref{fig:qalong}}. The momentum $p_{1}$
is $l$-collinear, the momenta $p_{2}$ and $p_{3}$ are $r$-collinear in
accordance with the classification of Table~\ref{tab:counting}. The
corresponding amplitudes for the diagrams Figs.\,(\ref{fig:qalong}a)
and (\ref{fig:qalong}b), which contribute to $G\left( \tau\right)$,
have the following form:
\begin{align}
\begin{split}
\bar{u}\left(  p_{2}\right)  \hat{V}_{\left(  a\right)  }^{\mu,a}v\left(
p_{1}\right)   &  =-g_{s}t^{a}\bar{u}\left(  p_{2}\right)  \gamma^{\mu}%
\frac{\not p_{2}+\not p_{3}}{\left(  p_{2}+p_{3}\right)  ^{2}+i0}\left(
\frac{\not n}{2}-\frac{\not n_{+}}{2}\right)  v\left(  p_{1}\right)  .\\
\bar{u}\left(  p_{2}\right)  \hat{V}_{\left(  b\right)  }^{\mu,a}v\left(
p_{1}\right)   &  =\,-g_{s}t^{a}\bar{u}\left(  p_{2}\right)  \left(
\frac{\not n}{2}-\frac{\not n_{+}}{2}\right)  \frac{-(\not p_{1}%
+\not p_{3})}{\left(  p_{1}+p_{3}\right)  ^{2}+i0}\,\gamma^{\mu}v\left(
p_{1}\right)  ,\label{eq:OffShellAmps}%
\end{split}
\end{align}
respectively. The sum of the amplitudes (\ref{eq:OffShellAmps}) has the
following $\lambda$-expansion:
\begin{multline}
\bar{u}\left(  p_{2}\right)  \left[  \hat{V}_{\left(  a\right)  }%
^{\mu,a}+\hat{V}_{\left(  b\right)  }^{\mu,a}\right]
v(p_{1})\\
=\bar{u}\left( p_{2}\right) \left( \frac{2g_{s}t^{a}}{Q}\right)  \left(
\,\gamma_{\perp}^{\mu}-\frac{n^{\mu }\not p_{3\perp}}{p_{3}\cdot
n+i0}\right)  v\left(  p_{1}\right)  \left[ 1+\frac{\left(
p_{2}+p_{3}\right)  ^{2}}{Q^{2}}\right]  +O\left(  \lambda ^{3}\right).
\end{multline}
Using the light-cone gauge, such that the gluon propagator has the form:%
\begin{equation}
\int\mathrm{d}^{D}x\,e^{-ip_{3}\cdot x}\left\langle 0\left\vert \mathrm{T}%
A_{a}^{\mu}\left(  x\right)  A_{b}^{\nu}\left(  0\right)  \right\vert
0\right\rangle =i\,\frac{\delta_{ab}}{p_{3}^{2}+i0}\left(  -g^{\mu\nu}%
+\frac{p_{3}^{\mu}n^{\nu}+p_{3}^{\nu}n^{\mu}}{p_{3}\cdot n+i0}\right)
,\label{eq:LCgauge}%
\end{equation}
and neglecting the power suppressed term $\left(  p_{2}+p_{3}\right)
^{2}/Q^{2}\sim \lambda^2$, we find that the following effective vertex
\begin{equation}
\hat{V}_{\mathrm{eff}}^{\mu,a}=\hat{V}_{\left(  a\right)  }^{\mu,a}+\hat
{V}_{\left(  b\right)  }^{\mu,a}=\frac{2g_{s}}{Q}t^{a}\,\frac{\not %
n\not n_{+}}{4}\gamma^{\mu}\frac{\not n\not n_{+}}{4}\label{eq:EffVert}%
\end{equation}
gives the leading contribution to $G\left(  \tau\right)  $. The $r$-
and $l$-collinear regions give equal contributions, and hence we find:
\begin{align}
\begin{split}
G^{(0)}\left(  \tau\right)  & =\frac{16\pi g_{s}^{2}C_{\mathrm{F}}}{Q^{4}}
\int\sum_{\sigma}\left\vert \bar{u}(p_{2})  \frac{\not %
n\not n_{+}}{4}\not \varepsilon^{\star}(p_{3})\frac{\not %
n\not n_{+}}{4}\,v(  p_{1})  \right\vert ^{2}\Theta\left(  \tau
Q^{2}-p_{\mathrm{R}}^{2}\right)  \mathrm{d}\rho_{3}\\
&=\,\frac{\alpha
_{s}C_{\mathrm{F}}}{\pi}\,\tau+O(\tau^{2}),\label{eq:Rtree}%
\end{split}
\end{align}
where $p_{\mathrm{R}}^{2}=(p_{2}+p_{3})^{2}$, $\varepsilon\left(
p_{3}\right) $ is the gluon polarization 4-vector $\varepsilon\left(
p_{3}\right) \cdot p_{3}=\varepsilon\left(  p_{3}\right) \cdot n=0$,
and $\mathrm{d}\rho_{3}$ is the element of the three particle phase
space (\ref{eq:PhSp}).

The expression (\ref{eq:EffVert}) remains valid in the light-cone gauge
(\ref{eq:LCgauge}) even when the particles are off shell ($p_{1}^{2}\sim p_{2}^{2}\sim p_{3}^{2}%
\sim\lambda^{2}Q^{2}$), i.e., it appears as the internal part of the
expansion of amplitudes with more than three collinear particles in the
final state\footnote{See Ref.\cite{Kirilin2010} for details.}.

The perturbative correction to $G^{(0)}\left(  \tau\right)  $ has the form \cite{Kirilin2010}:%
\begin{equation}
G\left(  \tau\right)  =G^{(0)}\left(  \tau\right)  \left[  1+\frac{\alpha_{s}%
}{4\pi}G^{(1)}\left(  \tau\right)  \right]  ,\label{eq:GPertResult}
\end{equation}
where $G^{(1)}\left(  \tau\right)  $ contains hard, $r$-collinear,
$l$-collinear and soft contributions:%
\begin{equation}
G^{(1)}\left(  \tau\right)  =-\frac{\beta_{0}}{\epsilon}+G_{\mathrm{hard}%
}^{(1)}\left(  Q^{2},\mu^{2}\right)  +G_{l-\operatorname{col}}^{(1)}\left(
\tau Q^{2},\mu^{2}\right)  +G_{r-\operatorname{col}}^{(1)}\left(  \tau
Q^{2},\mu^{2}\right)  +G_{\mathrm{soft}}^{(1)}\left(  \tau^{2}Q^{2},\mu
^{2}\right)  .
\end{equation}
where the term $-\beta_{0}/\epsilon$ appears after $\alpha_{s}$
renormalization in the vertex (\ref{eq:EffVert}). The hard contribution
consists only of virtual corrections to the amplitudes
(Fig.\ref{fig:qalong}), where the loop momentum is hard (Table \ref{tab:counting}):%
\begin{align}
G_{\mathrm{hard}}^{(1)}\left(  Q^{2},\mu^{2}\right)   &  =\left(  \frac{Q^{2}%
}{\mu^{2}}\right)  ^{-\epsilon}\left\{  -\frac{4C_{\mathrm{F}}}{\epsilon^{2}%
}+\frac{1}{\epsilon}\left[  \left(  \frac{2\pi^{2}}{3}-4\right)
C_{\mathrm{A}}+\left(  6-\frac{4\pi^{2}}{3}\right)  C_{\mathrm{F}}\right]
\right. \nonumber\\
&  \left.  +C_{\mathrm{A}}\left(  -16+\frac{2\pi^{2}}{3}+16\zeta(3)\right)
+C_{\mathrm{F}}\left(  34+\pi^{2}-32\zeta(3)\right)  +O\left(  \epsilon
\right)  \right\}  . \label{eq:GHard}%
\end{align}
As we shall see later, the hard contribution is by construction the
matching coefficient of the weak current to the effective three-body
operator, whose leading matrix element contains the vertex
(\ref{eq:EffVert}).

In contrast to Eq.(\ref{eq:R2Col}), $G_{l-\operatorname{col}}^{(1)}$
and $G_{r-\operatorname{col}}^{(1)}$ contributions are different,
because the leading amplitude contains two $r$-collinear particles and
only one $l$-collinear particle. Additional $l$-collinear particle
generates the same correction as Eq.(\ref{eq:R2Col}) for $F\left(
\tau\right)$, except for an additional integration over the invariant
mass $p_{\mathrm{R}}^{2}$ of the $r$-collinear particles:
\begin{align}
G_{l-\operatorname{col}}^{(1)}\left(  \tau Q^{2},\mu^{2}\right) & =\frac{4\pi
}{\alpha_{s}\tau}\int_{0}^{\tau Q^{2}}F^{l-\operatorname{col}}\left(  \tau-\frac{p_{\mathrm{R}}^{2}}{Q^{2}}\right) \frac{\mathrm{d} p_{\mathrm{R}}^{2}}{Q^{2}}
\notag \\ &=\left(  \frac{\tau Q^{2}}{\mu^{2}}\right)
^{-\epsilon}C_{\mathrm{F}}\left\{  \frac{4}{\epsilon^{2}}+\frac{7}{\epsilon
}+14-\pi^{2}+O\left(  \epsilon\right)  \right\}  .
\end{align}
The $r$-collinear contribution includes virtual corrections, so that
the loop momentum is implied to be $r$-collinear
(Table~\ref{tab:counting}), as well as
real radiation of three $r$-collinear and one $l$-collinear particles:%
\begin{multline}
G_{r-\operatorname{col}}^{(1)}\left(  \tau Q^{2},\mu^{2}\right) =\left(
\frac{\tau Q^{2}}{\mu^{2}}\right)  ^{-\epsilon}\left\{  \frac{4C_{\mathrm{F}}%
}{\epsilon^{2}}+\frac{1}{\epsilon}\left[  \left(  \frac{23}{3}-\frac{2\pi^{2}%
}{3}\right)  C_{\mathrm{A}}+\left(  \frac{4\pi^{2}}{3}-5\right)
C_{\mathrm{F}}-\frac{4N_{\mathrm{f}}T_{\mathrm{F}}}{3}\right]  \right. \\
+\left.  C_{\mathrm{A}}\left(  \frac{641}{18}-\frac{2\pi^{2}}{3}%
-22\zeta(3)\right)  +C_{\mathrm{F}}\left(  -\frac{95}{2}-\frac{\pi^{2}}%
{3}+44\zeta(3)\right)
-\frac{50}{9}\,N_{\mathrm{f}}T_{\mathrm{F}}+O\left( \epsilon\right)
\right\}  .
\end{multline}
The soft radiation also has the same form as Eq.(\ref{eq:R2Soft}) for
$F\left(\tau\right)$, but with an additional integration over the
invariant mass $p_{\mathrm{R}}^{2}$ of the $r$-collinear particles:
\begin{align}
G_{\mathrm{soft}}^{(1)}\left(  \tau^{2}Q^{2},\mu^{2}\right)  & = \frac{4\pi}
{\alpha_{s}\tau}\int_{0}^{\tau Q^{2}}F^{\mathrm{soft}}\left(\tau-\frac{p_{\mathrm{R}}^{2}}{Q^{2}}\right) \frac{\mathrm{d} p_{\mathrm{R}}^{2}}{Q^{2}}
\notag \\ &=\left(
\frac{\tau^{2}Q^{2}}{\mu^{2}}\right) ^{-\epsilon}C_{\mathrm{F}}\left\{
-\frac{4}{\epsilon^{2}}-\frac{8}{\epsilon
}+\frac{\pi^{2}}{3}-16\right\}  .
\end{align}

The singularities with respect to $\epsilon$ and the
$\mu^{2}$-dependence drop
out of the sum of all the contributions:%
\begin{align}
\begin{split}
G^{(1)}\left(  \tau\right) =&-4C_{\mathrm{F}}\ln^{2}\frac{1}{\tau}%
+\ln\frac{1}{\tau}\left[  C_{\mathrm{F}}\left(  \frac{4\pi^{2}}{3}-14\right)
+C_{\mathrm{A}}\left(  \frac{23}{3}-\frac{2\pi^{2}}{3}\right)  -\frac{4}%
{3}T_{\mathrm{F}}N_{\mathrm{f}}\right] \label{eq:R31}\\
&  +C_{\mathrm{F}}\left[  -\frac{31}{2}+12\zeta(3)\right]  +C_{\mathrm{A}%
}\left[  \frac{353}{18}-6\zeta(3)\right]  -\frac{50}{9}T_{\mathrm{F}%
}N_{\mathrm{f}}.
\end{split}
\end{align}

It is interesting to compare this correction with that of the leading
thrust distribution (\ref{eq:R2pert}). The double logarithmic
corrections are equal. It has a simple physical interpretation:
resummation of double logarithms results in a statistical factor
responsible for excluding some part of radiation which is out of the
two-jet region, and hance this factor is insensitive to the jet
structure. However, the single logarithms in Eq.(\ref{eq:R2pert}) and
Eq.(\ref{eq:R31}) are different. In the case of $F\left(  \tau\right)
$, it gives a positive contribution because
$6C_{\mathrm{F}}\ln\tau^{-1}>0$, but the correction to the $G(\tau)$
distribution (\ref{eq:R31}) contains the logarithm $\ln\tau^{-1}$\ with
a negative coefficient:
\begin{equation}
C_{\mathrm{F}}\left(  \frac{4\pi^{2}}{3}-14\right)  +C_{\mathrm{A}}\left(
\frac{23}{3}-\frac{2\pi^{2}}{3}\right)  -\frac{4}{3}T_{\mathrm{F}%
}N_{\mathrm{f}}=1-\frac{2\pi^{2}}{9}<0.
\end{equation}
The nonlogarithmic correction to $G^{\left(  1\right)  }\left(
\tau\right)  $ is positive and larger than $C_{\mathrm{F}}\left(
2\pi^{2}/3-2\right) $ in Eq.(\ref{eq:R2pert}). We will find in section
\ref{sec:resum} that these differences affect the event shape
distributions $F(\tau)$ and $G(\tau)$ after resummation of large
logarithmic corrections.

\section{Factorization formulae\label{sec:form}}

In this section we give the heuristic derivation of the factorization
formula for $G(\tau)$, which allows us to sum all big logarithmic
corrections to the NLL level of accuracy.

We first note that power counting rules imply a hierarchy of the
components of integration momenta and hence a hierarchy of the
components of fields. Instead of an expansion of integrands obtained in
a full QFT, it is sometimes possible to introduce an effective theory
such that its Feynman rules reproduce the expanded fundamental
amplitudes. Field modes corresponding to different regions are
associated with different fields of various effective theories.
Although, in perturbative calculations, the effective theory framework
does not provide new information, it turns out to be extremely
efficient when one needs to establish factorization formulae and
evolution equations for resummation of large logarithms. A lively
presentation of these ideas is provided e.g. in
Ref.\cite{Bauer:2008dt}, where, starting from the minimal set of
assumptions about hadronic final states, the authors use the
\textit{soft collinear effective theory} (SCET) \cite{Bauer2001,
Beneke2002} to derive the factorization formula for the distributions
of a large class of infrared safe observables (angularities).

In order to construct the SCET Lagrangian for collinear quarks
interacting with collinear and soft gluons, one has to split the quark
field $\psi$
using the projectors (\ref{eq:Pdef}):%
\begin{equation}
\psi=\xi_{n}+\eta_{n},\qquad\xi_{n}=\hat{P}_{\left(  +\right)  }\psi,\qquad
\eta_{n}=\hat{P}_{\left(  -\right)  }\psi,
\end{equation}
and split the gluon field into the collinear and soft parts:
\begin{equation}
A=A_{c,n}+A_{s}.
\end{equation}
The key assumption is that all the collinear field operators $\xi_{n}$,
$\eta_{n}$ and $A_{c,n}$, $A_{s}$ generate field modes with momenta
obeying the power counting rules listed in Table~\ref{tab:counting}.
Here collinear quanta corresponding to $A_{c,n}$ are $l$-collinear and
soft quanta $A_{s}$ have soft momenta. This assumption implies that we
restrict our consideration to the processes which can be described by
the QCD fields that are smooth in certain directions. The power
counting rules of Table \ref{tab:counting} lead to the following
hierarchy of the components of
fields:%
\begin{align}
\xi_{n} &  \sim\lambda, & \eta_{n} &  \sim\lambda^{2}, & A_{s}  &  \sim\lambda
^{2}\nonumber\\
n_{+}\cdot A_{c,n}  &  \sim1, & A_{c,n\,\perp}  &  \sim\lambda, & n\cdot
A_{c,n}  &  \sim\lambda^{2}, \label{eq:HeirFields}%
\end{align}
where the Sudakov decomposition of the collinear field $A_{c,n}$ is
used:
\begin{equation}
A_{c,n}^{\mu} = \left(n_{+} \cdot A_{c,n}\right)\frac{n^{\mu}}{2} + \left(n \cdot A_{c,n}\right)\frac{n_{+}^{\mu}}{2} +A_{c,n\perp}.
\end{equation}
Using the estimations (\ref{eq:HeirFields}), one can expand the QCD
Lagrangian to all orders in $\lambda$. This expansion truncated to some
order in such a way to preserve the invariance under the homogeneous
gauge transformations \cite{Beneke2003} corresponds to the SCET
Lagrangian \cite{Bauer2001, Beneke2002}.

Here, we outline the result of Ref.\cite{Bauer:2008dt} for the thrust
distribution:
\begin{equation}
F\left(  \tau\right)  =H\left(  Q^{2},\mu^{2}\right)  \int\mathrm{d}%
p_{\mathrm{L}}^{2}\mathrm{d}p_{\mathrm{R}}^{2}\mathrm{d}k\,J\left(
p_{\mathrm{L}}^{2},\mu^{2}\right)  J\left(  p_{\mathrm{R}}^{2},\mu^{2}\right)
S_{T}\left(  k,\mu^{2}\right)  \Theta\left(  Q^{2}\tau-p_{\mathrm{L}}%
^{2}-p_{\mathrm{R}}^{2}-Qk\right)  . \label{eq:ThrustMaster}%
\end{equation}
$H\left(  Q^{2},\mu^{2}\right)  $ is the hard function or the square of
the usual on-shell QCD Sudakov form factor. $J\left(
p^{2},\mu^{2}\right)  $
is the jet function:%
\begin{multline}
J\left(  p^{2},\mu^{2}\right)  =\frac{1}{\left(  p\cdot n_{+}\right)  N_{c}%
}\\
\times\frac{1}{2\pi}\operatorname{Im}\left[  i\int\mathrm{d}^{4}xe^{-ipx}%
\left\langle 0\left\vert \mathrm{T}\left\{ \bar{\xi}_{n}^{\prime}\left(
x\right)  W_{n}\left(  x\right) \frac{\not   n _{+}}{2}W_{n}^{\dagger
}\left(  0\right) \xi_{n}^{\prime}\left(  0\right)  \right\}
\right\vert
0\right\rangle \right]  , \label{eq:JetFunction}%
\end{multline}
that is, up to an overall factor, the imaginary part of the QCD quark
propagator in the light-cone gauge ($\xi'_{n}$ and and Wilson line
$W_{n}$ are defined below in Eq.(\ref{eq:GaugeFields})). The soft
factor $S_{T}\left( k,\mu^{2}\right)  $ is defined as follows:
\begin{equation}
S_{T}\left(  k,\mu^{2}\right)  =\sum_{X}\left\vert \left\langle X\left\vert
Y_{n_{+}}^{\dagger}Y_{n}\right\vert 0\right\rangle \right\vert ^{2}%
\delta\left(  k-n\cdot p_{X_{\mathrm{L}}}-n_{+}\cdot p_{X_{\mathrm{R}}%
}\right)  . \label{eq:SoftFactor}%
\end{equation}
$W$ and $Y$ are collinear and soft Wilson lines, respectively:
\begin{align}
W_{n}\left(  x\right) & =\mathrm{P}\exp\left[  ig_{s}\int_{-\infty}^{0
}\mathrm{d}s\,n_{+}\cdot A_{c}^{\prime}\left(  x+sn_{+}\right)  \right]
,\\
Y_{n}\left(  x\right) & =\mathrm{P}\exp\left[  ig_{s}\int_{-\infty}^{0}\mathrm{d}s\,n\cdot A_{s}^{\prime}\left(x+sn\right)  \right] .
\end{align}

In the effective theory framework the vertex (\ref{eq:EffVert})
corresponds to the following operator:
\begin{equation}\label{eq:O3LC}
\mathcal{O}_{3}   =\mathcal{O}_{3R}+\mathcal{O}_{3L},\qquad
\mathcal{O}_{3R}   =2g_{s}\bar{\xi}_{n_{_+}}\not   A_{\perp,n_{_+}}\,\xi
_{n},\qquad\mathcal{O}_{3L}=2g_{s}\bar{\xi}_{n_{_+}}\not   A_{\perp,n}\,\xi_{n},
\end{equation}
where $\xi_{n_{i}}$ and $A_{\perp n_{i}}$ are fields from different
SCET copies in the light-cone gauges of the type of
Eq.(\ref{eq:LCgauge}) with the light-like vector $n_{i}$. In order to
introduce the gauge invariant operators, one should replace the fields
entering the operators (\ref{eq:O3LC}) to $\xi^{\prime}$ and
$A_{\perp}^{\prime}$ in an arbitrary gauge, using the following
relations:
\begin{align}
\begin{split}
\xi &=YW^{\dagger}\xi^{\prime},\qquad g_{s}A_{c\perp}=Y\left(  W^{\dagger}iD_{c\perp}^{\prime}W-i\partial_{\perp}\right)  Y^{\dagger},\label{eq:GaugeFields}\\
D_{c\perp}^{\prime} &=\partial_{\perp}-igA^{\prime}_{c\perp}
\end{split}
\end{align}
where fields without primes are in the corresponding light-cone gauge.
Including a soft Wilson line $Y$ in the definitions
(\ref{eq:GaugeFields}) allows one to decouple soft and collinear
degrees of freedom in the leading order SCET Lagrangian
\cite{Bauer2002,Beneke2003}. Using the expressions
(\ref{eq:GaugeFields})
yields the following operators%
\begin{equation}
\mathcal{O}_{3}=2g_{s}\bar{\xi}_{n_{+}}^{\prime}\widetilde{\not   A }%
_{\perp,n_{+}}W_{n_{+}}Y_{n_{+}}^{\dagger}Y_{n}W_{n}^{\dagger}\xi_{n}^{\prime
}+2g_{s}\bar{\xi}_{n_{+}}^{\prime}W_{n_{+}}Y_{n_{+}}^{\dagger}Y_{n}%
W_{n}\widetilde{\not   A }_{\perp,n}\xi_{n}^{\prime}, \label{eq:O3}%
\end{equation}
where%
\begin{equation}
\tilde{A}_{\perp,n_{i}}=A_{\perp,n_{i}}^{\prime}-\frac{i}{g_{s}}W_{n_{i}}\left[  \partial_{\perp
},W^{\dagger}_{n_{i}}\right]  .
\end{equation}
The operator (\ref{eq:O3}) is in fact the operator $\mathcal{O}_{3}$
derived in Ref.\cite{Bauer:2006mk} taken in the limit:
$n_{q}\rightarrow n_{+},$ $n_{\bar{q}}\rightarrow n$ and
$n_{g}\rightarrow n$ or $n_{g}\rightarrow n_{+}$. Integration over hard
modes gives the matching coefficient $C_{H}$ of the QCD operator
$\hat{J}_{\parallel}$ (\ref{eq:Jparallel}) onto the SCET operator
$\mathcal{O}_{3}$ (\ref{eq:O3})
\begin{equation}\label{eq:CH}
\hat{J}_{\parallel} \to C_{H}\left(Q^{2},\mu^{2}\right)\mathcal{O}_{3}.
\end{equation}

The important point about the operator (\ref{eq:O3}) is that it is a
local product of the $r$-, $l$-collinear and soft SCET operators.
According to Ref.\cite{Bauer:2008dt}, this feature is the only
requirement to establish a factorization formula for an angularity
distribution. For the operator (\ref{eq:O3}), the thrust distribution
takes the form:
\begin{multline}
G\left(  \tau\right)  =2\,H_{3}\left(  Q^{2},\mu^{2}\right)  \int
\mathrm{d}p_{\mathrm{L}}^{2}\mathrm{d}p_{\mathrm{R}}^{2}\mathrm{d}%
k\,\Sigma_{\perp}\left(  p_{\mathrm{R}}^{2},\mu^{2}\right)  J\left(
p_{\mathrm{L}}^{2},\mu^{2}\right) \\
\times S_{T}\left( k,\mu^{2}\right) \Theta\left(
Q^{2}\tau-p_{\mathrm{L}}^{2}-p_{\mathrm{R}}^{2}-Qk\right) ,
\label{eq:R3}%
\end{multline}
where $S_{T}\left(  k,\mu^{2}\right)  $ is the same soft factor as
defined in (\ref{eq:SoftFactor}), $J\left(
p_{\mathrm{L}}^{2},\mu^{2}\right)  $ is the jet function defined in
(\ref{eq:JetFunction}), and $H_{3}=\left\vert C_{H}\right\vert ^{2}$.
The new object in the formula (\ref{eq:R3}) is $\Sigma_{\perp}\left(
p_{\mathrm{R}}^{2},\mu^{2}\right)  $:
\begin{multline}
\Sigma_{\perp}\left(  p^{2},\mu^{2}\right)  =\frac{g_{s}^{2}}{\left(
p\cdot n\right)
Q^{2}N_{\mathrm{c}}}\frac{1}{\pi}\\
\times\operatorname{Im}\left[
i\int\mathrm{d}^{\mathcal{D}}xe^{-ipx}\left\langle 0\left\vert \mathrm{T}%
\left\{  \left(  \bar{\xi}_{n_{+}}^{\prime}\widetilde{\not   A }%
_{\perp,n_{+}}W_{n_{+}}\right)  \left(  x\right)  \frac{\not   n
}{2}\left( W_{n_{+}}^{\dagger}\widetilde{\not   A
}_{\perp,n_{+}}\xi_{n_{+}}^{\prime }\right)  \left(  0\right)  \right\}
\right\vert 0\right\rangle \right]  ,
\label{eq:Sigma}%
\end{multline}
which can be considered as the imaginary part of the quark\
\textquotedblleft transverse\textquotedblright\ self energy projected
onto $\not  n $. In contrast to the jet function or the soft factor,
whose leading expressions are $\delta$-functions, the tree level
expression for $\Sigma_{\perp
}\left(  p^{2},\mu^{2}\right)  $ is a smooth function:%
\begin{equation}
\Sigma_{\perp}^{\left(  0\right)  }\left(  p_{\mathrm{R}}^{2},\mu^{2}\right)
=\frac{\alpha_{s}\left(  \mu^{2}\right)  C_{\mathrm{F}}}{4\pi Q^{2}}\left(
\frac{p_{\mathrm{R}}^{2}}{4\pi\mu^{2}}\right)  ^{\mathcal{D}/2-2}\frac
{2\Gamma\left(  \mathcal{D}/2\right)  }{\Gamma\left(  \mathcal{D}-2\right)
}.
\end{equation}

The most important property of the operator (\ref{eq:O3}) is that it is
a local product of the collinear fields. The reason is that dominant
underlying process shown in \Fig{\ref{fig:inter2}} does not depend on
the kinematics of the final state since the energy of the intermediate
state
$E_{2}=Q+\left\vert \mathbf{p}\right\vert +\left\vert \mathbf{p}%
_{2}\right\vert +\left\vert \mathbf{k}\right\vert $ is equal to $2Q$
due to energy conservation for the final state. It results in locality
of the vertex (\ref{eq:EffVert}) for the three-particle interactions.
The essential consequence of the locality is that soft gluons are
radiated off by just like the two-prong QCD antenna
(\ref{eq:SoftFactor}) for the leading contribution to the structure
function $F$; see Eq.(\ref{eq:ThrustMaster}). An intuitive explanation
of this fact is that soft gluons are coherently radiated off all
collinear emitters and the factorized amplitude of the soft radiation
does not depend on the fraction of the collinear momentum carried by a
particular emitter. For example, using a noncovariant gauge, one can
show that the amplitude of soft radiation off the $r$-collinear pair
depicted in \Fig{\ref{fig:qalong}} is equivalent to the corresponding
amplitude for a single quark\footnote{In fact, the same argument is the
same as that for explaining the \textit{angular ordering} in QCD
sequential branching process (Ref.\cite{Ermolaev:1981cm,
Mueller:1981ex, Fadin1983}).}:
\begin{equation}
t^{a}t^{b}\,\frac{e\left(  k\right)  \cdot n_{+}}{k\cdot n_{+}}-if^{abc}%
t^{c}\,\frac{e\left(  k\right)  \cdot n_{+}}{k\cdot n_{+}}=t^{b}t^{a}%
\,\frac{e\left(  k\right)  \cdot n_{+}}{k\cdot n_{+}}.
\end{equation}

All those remarkable properties of the structure function $G$ such as
the locality of the corresponding effective operator and the universal
soft factor can be hardly generalized to the difference $F-K$, where
the third topology (Fig.\ref{fig:topo}c) accounts for the leading
contribution (\ref{eq:SmallTauExpFK}). In this case the soft
radiation undoubtedly differs from that for the topologies (Fig.\ref{fig:topo}%
a,\thinspace b) because one hemisphere contains the single high energy
gluon only. Moreover, the part of the cross section contributing to
$F-K$ is singular in the region where the energy of the quark or
antiquark tends to zero. This
singularity, which is quite similar to that in the $\gamma^{\star}%
\gamma\rightarrow\pi^{0}$ process, leads to an additional $\ln\tau$ in
the difference $F-K$ in Eq.(\ref{eq:SmallTauExpFK}). It indicates that
the effective operator for $F-K$ should be nonlocal in light-cone
directions and the factorized objects should be different from those
used in Eq.(\ref{eq:R3}). We will discuss $F-K$ contribution to single
flavor tag measurements briefly in section \ref{sec:fb}.

\section{Resummation of large logarithms\label{sec:resum}}

In order to perform integration in the factorization formula (\ref{eq:R3}) we
take the Laplace transform of each function entering Eq.(\ref{eq:R3}) except
for the hard coefficient function $H_{3}\left(  Q^{2},\mu^{2}\right)  $,%
\begin{equation}
G\left(  \tau\right)  =2\,H_{3}\left(  Q^{2},\mu^{2}\right)  \frac{1}{2\pi
i}\int_{C}\frac{\mathrm{d}\nu}{\nu}\,e^{\nu Q^{2}\tau}\,\tilde{\Sigma}_{\perp}\left(  sQ^{2}%
,\mu^{2}\right)  j\left(  sQ^{2},\mu^{2}\right)  s_{T}\left(  sQ,\mu
^{2}\right)  , \label{eq:R3Laplace}%
\end{equation}
where $s=1/\left(  \nu Q^{2}e^{\gamma_{E}}\right)  $ and%
\begin{align}
j\left(  sQ^{2},\mu^{2}\right)   &  \equiv\int_{0}^{\infty}\mathrm{d}%
p^{2}e^{-\nu p^{2}}J\left(  p^{2},\mu^{2}\right)  ,\qquad s_{T}\left(
sQ,\mu^{2}\right)  \equiv\int_{0}^{\infty}\mathrm{d}k\,e^{-\nu Qk}S_{T}\left(
k,\mu^{2}\right)  ,\notag\\
\tilde{\Sigma}_{\perp}\left(  sQ^{2},\mu^{2}\right)   &  \equiv\int
_{0}^{\infty}\mathrm{d}p_{\mathrm{R}}^{2}e^{-\nu p_{\mathrm{R}}^{2}}%
\,\Sigma_{\perp}\left(  p_{\mathrm{R}}^{2},\mu^{2}\right)  .
\end{align}
Now we use the fact that the expression (\ref{eq:R3Laplace}) does not
depend on $\mu^{2}$. We can exclude all collinear logarithms by setting
$\mu^{2}=\tau Q^{2}\sim\lambda^{2}$. In doing so,\ we can neglect all
higher-order corrections in $\tilde{\Sigma}_{\perp}\left(
sQ^{2},\mu^{2}\right)  $ and $j\left(  sQ^{2},\mu^{2}\right)  $ since
all of them contribute either to the N$^{2}$LL level, or to the
(pre-exponential) factor, which does not depend on $\tau$ and can be
found from the fixed-order result (\ref{eq:R31}). Therefore, we can
replace in Eq.(\ref{eq:R3Laplace}) the jet function $j\left(
sQ^{2},\mu^{2}\right) $ by unity and $\tilde{\Sigma}_{\perp}\left(
sQ^{2},\mu^{2}\right)  $ by $\tilde{\Sigma}_{\perp}\left(  sQ^{2},\tau
Q^2\right)$:
\begin{equation}
\tilde{\Sigma}_{\perp}\left(  sQ^{2},\tau Q^{2}\right)  =\int_{0}%
^{\infty}\mathrm{d}p_{\mathrm{R}}^{2}e^{-\nu p_{\mathrm{R}}^{2}}\,\left.
\Sigma_{\perp}^{(0)}\left(  p_{\mathrm{R}}^{2},\tau Q^{2}\right)  \right\vert
_{\mathcal{D}=4}=\frac{\alpha_{s}\left(  \tau Q^{2}\right)  C_{\mathrm{F}}%
}{2\pi}\frac{1}{\nu Q^{2}}.
\end{equation}
Thus, we obtain the following distribution:%
\begin{equation}
G\left(  \tau\right)  =\frac{2\alpha_{s}\left(  \tau Q^{2}\right)
C_{\mathrm{F}}}{\pi}\,H_{3}\left(  Q^{2},\tau Q^{2}\right)  \frac{1}{2\pi
i}\int_{C}\frac{\mathrm{d}\nu}{\nu^{2}Q^{2}}\,e^{\nu Q^{2}\tau} s_{T}\left(  sQ,\tau
Q^{2}\right)  . \label{eq:R3simpl}%
\end{equation}

Let us now consider the evolution equation for the hard
coefficient functions and the soft factor\cite{Becher2008}:%
\begin{align}
\frac{\mathrm{d}H_{i}\left(  Q^{2},\mu^{2}\right)  }{\mathrm{d}\ln\mu^{2}}  &
=\left\{  \Gamma_{\mathrm{cusp}}\left[  \alpha_{s}\left(  \mu^{2}\right)
\right]  \ln\frac{Q^{2}}{\mu^{2}}+\gamma^{\mathrm{H}_{i}}\left[  \alpha
_{s}\left(  \mu^{2}\right)  \right]  \right\}  H_{i}\left(  Q^{2},\mu
^{2}\right)  ,\nonumber\\
\frac{\mathrm{d}s_{T}\left(  sQ,\mu^{2}\right)  }{\mathrm{d}\ln\mu^{2}}  &
=\left\{  \Gamma_{\mathrm{cusp}}\left[  \alpha_{s}\left(  \mu^{2}\right)
\right]  \ln\frac{s^{2}Q^{2}}{\rho^{2}}-\gamma^{\mathrm{S}}\left[  \alpha
_{s}\left(  \mu^{2}\right)  \right]  \right\}  s_{T}\left(  sQ^{2},\mu
^{2}\right)  , \label{eq:EvEq}%
\end{align}
where%
\begin{align}
\Gamma_{\mathrm{cusp}}\left(  \alpha_{s}\right)   &  =\frac{\alpha_{s}}{4\pi
}\,\Gamma_{\left(  0\right)  }+\left(  \frac{\alpha_{s}}{4\pi}\right)
^{2}\,\Gamma_{\left(  1\right)  }+\ldots,\notag\\
\gamma^{i}\left(  \alpha_{s}\right)   &  =\frac{\alpha_{s}}{4\pi}%
\,\gamma_{\left(  0\right)  }^{i}+\left(  \frac{\alpha_{s}}{4\pi}\right)
^{2}\,\gamma_{\left(  1\right)  }^{i}+\ldots.
\end{align}
In the NLL accuracy, one needs the following expressions: two-loop
$\,\Gamma_{\mathrm{cusp}}$, two-loop $\alpha_{s}\left(  \mu^{2}\right)
$, one-loop $\gamma^{i}$:
\begin{align}
\Gamma_{\left(  0\right)  }  &  =4C_{\mathrm{F}},\qquad\Gamma_{\left(
1\right)  }=\frac{4C_{\mathrm{F}}}{9}\left[  C_{\mathrm{A}}\left(  67-3\pi
^{2}\right)  -20T_{\mathrm{F}}N_{\mathrm{f}}\right]  ,\notag\\
\gamma_{\left(  0\right)  }^{H_{2}}  &  =-6C_{\mathrm{F}},\qquad
\gamma_{\left(  0\right)  }^{H_{3}}=\left(  \frac{2\pi^{2}}{3}-4\right)
C_{\mathrm{A}}+\left(  6-\frac{4\pi^{2}}{3}\right)  C_{\mathrm{F}},
\end{align}
$\Gamma_{\left(  1\right)  }$ is found in Ref.\cite{Kodaira1982},
$\gamma_{\left(  0\right)  }^{H_{2}}$ and $\gamma_{\left(  0\right)
}^{H_{3}}$ can be found from the expressions (\ref{eq:R2Hard}) and
(\ref{eq:GHard}), respectively. The initial conditions $H_{i}(Q^2,Q^2)$
and $s_{T}(sQ,sQ)$ for the equations (\ref{eq:EvEq}) contribute to the
pre-exponential factor.

Let us, for a moment, omit all pre-exponential factors and set $\mu^{2}=\tau
Q^{2}$, thus the solutions of the equations (\ref{eq:EvEq}) can be represented
as follows:%
\begin{align}
H_{i}\left(  Q^{2},\tau Q^{2}\right)   &  =\exp\left\{  \mathcal{F}_{H_{i}%
}\left[  L,\alpha_{s}\left(  Q^{2}\right)  \right]  \right\}  ,\notag\\
s_{T}\left(  sQ,\tau Q^{2}\right)   &  =\exp\left\{  \mathcal{F}_{s}\left[
L,\tilde{L},\alpha_{s}\left(  Q^{2}\right)  \right]  \right\}  ,
\end{align}
where%
\begin{align}
\mathcal{F}_{H_{i}}\left[  L,\alpha_{s}\left(  Q^{2}\right)  \right]   &
=-\int_{\tau Q^{2}}^{Q^{2}}\frac{\mathrm{d}\tilde{\mu}^{2}}{\tilde{\mu}^{2}}\left(
\Gamma_{\mathrm{cusp}}\left[  \alpha_{s}\left(  \tilde{\mu}^{2}\right)  \right]
\ln\frac{Q^{2}}{\tilde{\mu}^{2}}+\gamma^{\mathrm{H}_{i}}\left[  \alpha_{s}\left(
\tilde{\mu}^{2}\right)  \right]  \right)  ,\nonumber\\
\mathcal{F}_{s}\left[  L,\tilde{L},\alpha_{s}\left(  Q^{2}\right)  \right]
&  =\int_{s^{2}Q^{2}}^{\tau Q^{2}}\frac{\mathrm{d}\tilde{\mu}^{2}}{\tilde{\mu}^{2}}\left(
\Gamma_{\mathrm{cusp}}\left[  \alpha_{s}\left(  \tilde{\mu}^{2}\right)  \right]
\ln\frac{s^{2}Q^{2}}{\tilde{\mu}^{2}}-\gamma^{\mathrm{S}}\left[  \alpha_{s}\left(
\tilde{\mu}^{2}\right)  \right]  \right)  , \label{eq:Exponents}%
\end{align}
and%
\begin{equation}
L=\ln\frac{1}{\tau},\qquad\tilde{L}=\ln \frac{\tau}{s e^{\gamma_{E}}}  =\ln\left(  \tau\nu Q^{2}\right)  .
\end{equation}

The method of calculating the integral transform in (\ref{eq:R3simpl})
is developed in Ref.\cite{Catani1993}. It is based on the expansion of
the function $\mathcal{F}_{s}\left[  L,\tilde{L},\alpha_{s}\left(
Q^{2}\right) \right]  $ into the power series with respect to
$\tilde{L}$ and using the
simple formula%
\begin{equation}
\frac{1}{2\pi i}\int_{C}\mathrm{d}u\,\ln^{k}u\,e^{u-(1-g)\ln u}=\frac
{\mathrm{d}^{k}}{\mathrm{d}g^{k}}\frac{1}{\Gamma\left(  1-g\right)  }.
\end{equation}
Using this method, we obtain the following result for the resummed
distribution:%
\begin{equation}
G\left(  \tau\right)  =\left[  1+\mathcal{C}_{3}\alpha_{s}\left(
Q^{2}\right)  \right]  \,\frac{\alpha_{s}\left(  \tau Q^{2}\right)  \tau
C_{\mathrm{F}}}{\pi}\,\frac{\exp\left[  \mathcal{F}_{H_{3}}\left(
L,\alpha_{s}\right)  +\mathcal{F}_{s}\left(  L,0,\alpha_{s}\right)  \right]
}{\Gamma\left[  2-g\left(  L,\alpha_{s}\right)  \right]  }, \label{eq:R3resum}%
\end{equation}
where $\alpha_{s}=\alpha_{s}\left(  Q^{2}\right)  $ and%
\begin{equation}
g\left(  L,\alpha_{s}\right)  =\left.  \frac{\partial}{\partial\tilde{L}%
}\mathcal{F}_{s}\left(  L,\tilde{L},\alpha_{s}\right)  \right\vert _{\tilde
{L}=0}.
\end{equation}
In the expression (\ref{eq:R3resum}), we have restored the
pre-exponential factor $1+\alpha_{s}\left(  Q^{2}\right)
\mathcal{C}_{3}$, where
\begin{equation}
\mathcal{C}_{3}=C_{\mathrm{F}}\left[  -\frac{31}{2}+12\zeta(3)\right]
+C_{\mathrm{A}}\left[  \frac{353}{18}-6\zeta(3)\right]  -\frac{50}%
{9}T_{\mathrm{F}}N_{\mathrm{f}}.
\end{equation}
The corresponding expression for $F\left(  \tau\right)  $ has the form:%
\begin{equation}
F\left(  \tau\right)  =\left[  1+\mathcal{C}_{2}\alpha_{s}\right]
\,\frac{\exp\left[  \mathcal{F}_{H_{2}}\left(  L,\alpha_{s}\right)
+\mathcal{F}_{s}\left(  L,0,\alpha_{s}\right)  \right]  }{\Gamma\left[
1-g\left(  L,\alpha_{s}\right)  \right]  }, \label{eq:R2resum}%
\end{equation}
where%
\begin{equation}
\mathcal{C}_{2}=C_{\mathrm{F}}\left(  -2+\frac{2\pi^{2}}{3}\right)  .
\label{eq:C2}%
\end{equation}
The SCET result (\ref{eq:R2resum}) presented in Eq.(\ref{eq:R2resum})
coincides with the result of Ref.\,\cite{Catani1993}.

Now one can compare the leading thrust distribution (\ref{eq:R2resum}) and the
power suppressed one (\ref{eq:R3resum}):%
\begin{equation}
\frac{G\left(  \tau\right)  }{F\left(  \tau\right)  }=C_{\mathrm{F}}%
\,\frac{\alpha_{s}\left(  \tau Q^{2}\right)  }{\pi}\tau\,\left[  1+\alpha
_{s}\left(  Q^{2}\right)  \left(  \mathcal{C}_{3}-\mathcal{C}_{2}\right)
\right]  \frac{\exp\left\{  \mathcal{F}_{H_{3}}\left[  L,\alpha_{s}\left(
Q^{2}\right)  \right]  -\mathcal{F}_{H_{2}}\left[  L,\alpha_{s}\left(
Q^{2}\right)  \right]  \right\}  }{1-g\left[  L,\alpha_{s}\left(
Q^{2}\right)  \right]  }. \label{eq:Ratio}%
\end{equation}
Taking into account the explicit form of the exponents
(\ref{eq:Exponents}),
we obtain
\begin{equation}
\frac{G\left(  \tau\right)  }{F\left(  \tau\right)  }=G^{(0)}\left(
\tau\right)  \,e^{\omega\left(  \tau\right)  }, \label{eq:GtoF}%
\end{equation}
where%
\begin{equation}
\omega\left(  \tau\right)    =\frac{\gamma_{\left(  0\right)  }^{H_{3}%
}-\gamma_{\left(  0\right)  }^{H_{2}}-\beta_{0}}{\beta_{0}}\ln\left(
1-\lambda\right)  -\ln\left[  1-g\left(  L,\alpha_{s}\right)  \right]
+\alpha_{s}\left(  \mathcal{C}_{3}-\mathcal{C}_{2}\right)  ,
\end{equation}
with
\begin{equation}
g\left(  L,\alpha_{s}\right)  =\frac{2\Gamma_{0}}{\beta_{0}}\,\left[
\ln\left(  1-2\lambda\right)  -\ln\left(  1-\lambda\right)  \right]
,\qquad
\lambda =\frac{\beta_{0}\alpha_{s}}{4\pi}\ln\frac{1}{\tau}.
\end{equation}
Here the pre-exponential factor $1+\alpha_{s}\left(  Q^{2}\right)
\left( \mathcal{C}_{3}-\mathcal{C}_{2}\right)$ has also been
exponentiated to the NLL level of accuracy. Since the soft factor in
Eq.(\ref{eq:R3}) is the same as the one in Eq.(\ref{eq:ThrustMaster}),
it drops out of the ratio (\ref{eq:Ratio}) almost completely. The
resummation factor $\exp\left[\omega(\tau)\right]$ is shown in
\Fig{\ref{fig:expomega}}.

\FIGURE[t]{\epsfig{file=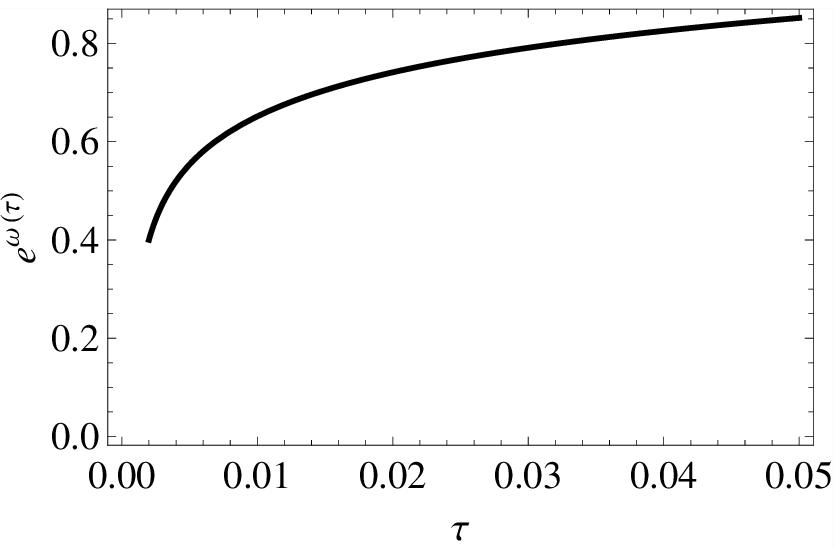,width=.4\textwidth}
\caption{Resummation factor as a function of the thrust boundary
$\tau>1-T$.\label{fig:expomega}}}

\FIGURE[b]{\epsfig{file=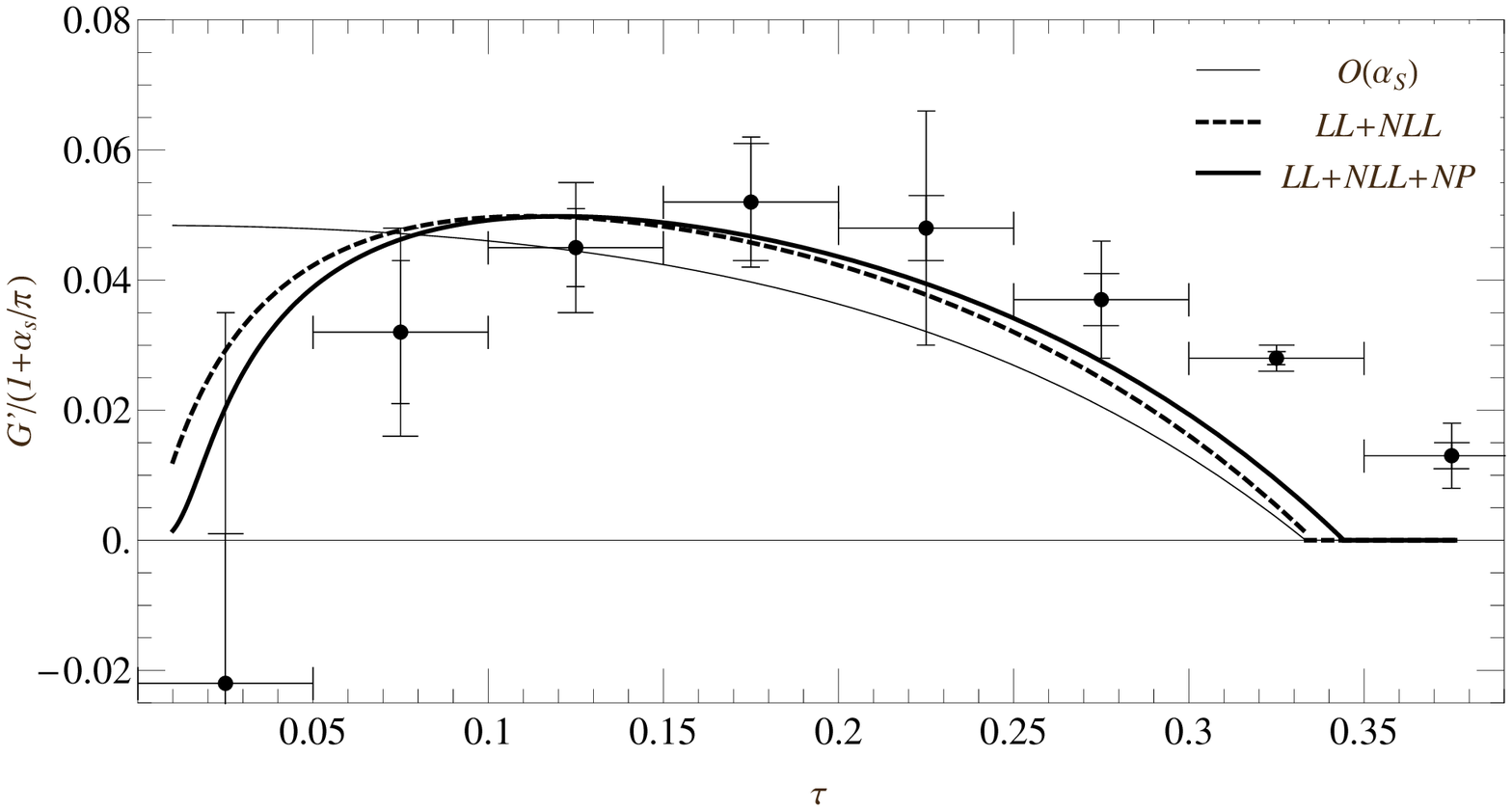,width=.7\textwidth}
\caption{Comparison of the OPAL data for the longitudinal thrust event
shape with the theoretical predictions $\mathrm{d}G/\mathrm{d}\tau$ at
$\tau=1-T$. The bars with short strokes represent systematic errors and
those with long strokes are statistical errors. The thin solid curve
shows $O\left(\alpha_{s}\right)$ perturbative result, the dashed curve
gives the LL+NLL prediction, while the solid curve is obtained after
convoluting with the non-perturbative shape function discussed in
section \ref{sec:np}.\label{fig:Opal}}}

Since the function $G(\tau)$ gives the shape of the jets which have
distinct angular distribution $\sim\sin^{2}\theta_{T}$, which is
different from that of $F(\tau)$, there is a possibility to measure
$G(\tau)$. Such analysis was performed by the OPAL collaboration
\cite{Abbiendi1998}. The comparison of theoretical predictions with the
OPAL data is presented in \Fig{\ref{fig:Opal}}, where three curves are
shown. The thin solid line corresponds to the perturbative result
(\ref{eq:G1}). As one can see from Eq.(\ref{eq:Rtree}),
$\mathrm{d}G^{(0)}/\mathrm{d}\tau$ tends to constant in the $\tau \to
0$ limit. The distribution improved by resummation (\ref{eq:R3resum})
is drawn by the dashed line. The solid line present the prediction with
nonperturbative effects as discussed below in Sect.\,\ref{sec:np}.

Since the result (\ref{eq:R3resum}) is valid in the region $\tau\ll1$,
we match the resummation factors with the perturbative result
(\ref{eq:G1}) so that all higher order corrections disappear when
$\tau$ tends to its maximal value for a three-jet configuration
$\tau_{\max}=1/3$. The lack of multiplicity for the perturbative result
(\ref{eq:G1}) explains why the data exceeds the prediction in the
region $\tau\gtrsim1/3$, while the poor accuracy of the data in the
region $\tau\ll1$ does not allow one to test the NLL effect specific
for $G\left(  \tau\right)  $ against the background of the common
Sudakov suppression.

The factorization scale $\mu^{2}$ presented in
Eqs.(\ref{eq:ThrustMaster}), (\ref{eq:R3}) formally separates collinear
\textit{intra}-jet radiation and soft \textit{inter}-jet radiation.
Since an infrared safe event shape does not depend on an explicit
definition of jet, the factorization formulae do not depend on this
scale. This fact is very helpful in deriving the simple representation
(\ref{eq:R3simpl}) for $G\left(  \tau\right)  $ where the role of the
collinear scale is reduced to the renormalization scale of $\alpha_{s}$
and determination of the argument of the hard logarithms in
Eq.(\ref{eq:Ratio}). It is worth noticing that we use the same
$\Gamma_{\mathrm{cusp}}$ in the evolution equations (\ref{eq:EvEq}) for
all $H_{i}$, although, we test it only in the leading order (see
Eq.(\ref{eq:GHard})).

\section{Nonperturbative correction\label{sec:np}}

For the total thrust distribution, the resummation of large logarithms
was performed to NLL accuracy in Ref.\cite{Catani1993} by means of the
NLL branching algorithm. The authors of Ref.\cite{Catani1993} notice
that in order to keep the algorithm in the perturbative regime, one has
to introduce the infrared regulator for the argument of the running
coupling constant. Alternatively, as it was demonstrated in
Refs.\cite{Catani:1991kz, Catani:1991bd, Dokshitzer1997}, one may take
into account non-perturbative (NP) effects by convoluting the resummed
perturbative expression with a phenomenological shape function $u\left(
\tau\right)$:
\begin{equation}
\sigma_{_\mathrm{NP+PT}}\left(  \tau\right)  =\int_{0}^{\tau Q/\Lambda}\mathrm{d}\tau^{\prime}%
\sigma_{_\mathrm{PT}}\left(  \tau-\frac{\Lambda}{Q}\,\tau^{\prime}\right)
u\left(  \tau^{\prime}\right),\label{eq:ConvNP}
\end{equation}
where $\Lambda$ is a phenomenological soft scale characterizing the
transition into the NP regime. Non-perturbative power corrections
generated by this shape function were the subject of intensive
experimental\cite{Dissertori:1999kj,Abreu:1999rc,Pfeifenschneider:1999rz,Biebel:2000zc,Acciarri:2000hm}
and theoretical \cite{Dokshitzer1995, Dokshitzer1996, Dokshitzer1997,
Korchemsky1999} studies. Since the NP corrections appear mostly due to
the radiation of soft partons, NP effects should firstly affect the
generalized soft factor:
\begin{equation}
S_{T}\left(\alpha,\beta;\mu^2\right)  =\sum_{X}\left\vert \left\langle X\left\vert
Y_{n_{+}}^{\dagger}Y_{n}\right\vert 0\right\rangle \right\vert ^{2}%
\delta\left(  \alpha-n\cdot p_{_{X_{\mathrm{L}}}}\right) \delta\left( \beta-n_{+}\cdot p_{_{X_{\mathrm{R}}}}\right).
\end{equation}
It implies that NP effects are universal for the distributions
containing the same factorized soft factor, as confirmed by  the
analysis performed in Ref.\cite{Korchemsky1998}. Since $F$ and $G$
structure functions contain the same soft factor, one can estimate the
influence of NP effects on the ratio $G/(F+G)$ (see
Eqs.(\ref{eq:ThrustMaster}) and (\ref{eq:R3})). A very simple but
reasonably good parametrization for the shape function was found in
Ref.\cite{Korchemsky1998}:
\begin{equation}
u_{_\mathrm{K}}\left(  x\right)  =\frac{2}{\Gamma\left(  3/2\right)
}\,x^{2}e^{-x^{2}},\qquad\Lambda=0.7\,\mathrm{GeV}.\label{eq:uK}%
\end{equation}
Using this parametrization, we find that the NP correction reduces to a
simple shift of the cross section even in a region $\tau>1-\langle
T\rangle=0.066$, so that the relative correction is
\begin{equation}
\frac{\left[G/(F+G)\right]_{\mathrm{PT+NP}}-\left[G/(F+G)\right]_{\mathrm{PT}}}{\left[G/(F+G)\right]_{\mathrm{PT}}}=-0.22\label{eq:NPEffect}
\end{equation}
for $\tau = 1-\langle T \rangle$ and increases for smaller $\tau$ (see
\Fig{\ref{fig:CCoef}} below).

Since the NP shape function (\ref{eq:uK}) is normalized to unity, the
NP correction to the structure function integrated over total domain,
i.e. $\int \mathrm{d}\tau \mathrm{d}G/\mathrm{d\tau}$, should vanish.
However, the convolution with $u_{\mathrm{K}}\left( x\right)$ shifts
the point where $\mathrm{d}G/\mathrm{d\tau}=0$. Thus, if we restrict
the integration to the true value of $\tau_{\max}=1/2$, then the NP
correction to $G(\tau_{\max})$ would be very small. This statement is
in agreement with the Monte-Carlo study of hadronization effects
performed in Ref.\,\cite{Abbaneo1998}, where some generators found no
correction and some generators found a small positive correction to the
inclusive combination $(F-2G)/(F+2G)$.

\section{Forward-backward asymmetry\label{sec:fb}}

In order to suppress QCD corrections to the forward-backward (FB)
asymmetry, one can reduce final state phase space to the two jet
region, thereby suppressing gluon radiation. It has been studied in
Ref.\cite{Abbaneo1998} how the experimental cuts bias the theoretical
corrections. The event shape can also been used to select the events.
Since the phase space of two $r$-collinear partons is of the order
$\tau$ in the $\tau\rightarrow0$ limit the corresponding correction
decreases with $\tau$ (see Refs.\cite{Lampe1993, Djouadi1995} and
\Fig{\ref{fig:CCoef}}).

We define the FB asymmetry, which depends on the maximal thrust value
($T<1-\tau$), as follows
\begin{equation}
A\left(  \tau\right)  =\frac{\int_{0}^{1}\mathrm{d}\cos\theta_{T}\,w\left(
\theta_{T},\tau\right)  -\int_{-1}^{1}\mathrm{d}\cos\theta_{T}\,w\left(
\theta_{T},\tau\right)}{\int_{-1}^{1}\mathrm{d}\cos\theta_{T}\,w\left(
\theta_{T},\tau\right)  }=A^{(0)}\frac{K\left(  \tau\right)  }{F\left(
\tau\right)  +G\left(  \tau\right)  }.
\end{equation}
Here
\begin{equation}
w\left(  \theta,\tau\right)  =\int_{1-\tau}^{1}\mathrm{d}T\,\frac
{\mathrm{d}\sigma}{\mathrm{d}\cos\theta_{T}\mathrm{d}T}, \label{eq:w}%
\end{equation}
with $\mathrm{d}\sigma\left(  \tau\right)  $ defined in
Eq.(\ref{eq:SigmaT}),
and $A^{(0)}$ is the tree-level asymmetry at $Q^{2}=M_{Z}^{2}$ :%
\begin{equation}
A^{(0)}=\frac{3}{4}\frac{2g_{al}g_{vl}}{\left(  g_{al}^{2}+g_{vl}^{2}\right)
}\frac{2g_{aq}g_{vq}}{\left(  g_{aq}^{2}+g_{vq}^{2}\right)  }.
\end{equation}
The effect of QCD radiative corrections can be characterized by the
following
quantity:%
\begin{align}
C\left(  \tau\right)   &  =1-\frac{A\left(  \tau\right)  }{A^{(0)}}=C^{\left(
F-K\right)  }\left(  \tau\right)  +C^{\left(  G\right)  }\left(  \tau\right)
,\label{eq:C}\\
C^{\left(  F-K\right)  }\left(  \tau\right)   &  =\frac{F\left(  \tau\right)
-K\left(  \tau\right)  }{F\left(  \tau\right)  +G\left(  \tau\right)
},\label{eq:CFK}\\
C^{\left(  G\right)  }\left(  \tau\right)   &  =\frac{G\left(  \tau\right)
}{F\left(  \tau\right)  +G\left(  \tau\right)  }. \label{eq:CG}%
\end{align}
The coefficients $C^{\left(  F-K\right)  }$ and $C^{\left(  G\right) }$
vanish if the free parton model relations (\ref{eq:FeqK}),
(\ref{eq:Geq0}) hold. The sum $F(\tau)+G(\tau)$ gives the integrated
thrust cross section.


We show in \Fig{\ref{fig:CCoef}} QCD predictions for the correction
factor $C\left(  \tau\right)  $ which relates as in Eq.(\ref{eq:C}) the
electroweak parameter and the observable FB asymmetry of $q\bar{q}$
jets whose thrust value is greater than $T>1-\tau$. \ Shown by thin
solid lines are the tree-level results for $C^{\left(  F-K\right)  }$
and $C^{\left(  G\right) } $. The correction factor $C^{\left(
F-K\right)  }$ arises from the three parton ($q\bar{q}g$) configuration
where both $q$ and $\bar{q}$ are in the same hemisphere along the
thrust axis, as illustrated in \Fig{\ref{fig:topo}}(c). Those events
cannot contribute to the P-odd function $K\left( \tau\right)  $ and the
observed asymmetry reduces by the factor $C^{\left(  F-K\right)  }$. As
one can see from Eq.\,(\ref{eq:SmallTauExpFK}), the derivative of the
function $F_{\mathrm{tree}}\left( \tau\right) -K_{\mathrm{tree}}\left(
\tau\right)  $ has logarithmic singularity in the two-jet limit
$T\rightarrow1$ (see Eq.\thinspace(\ref{eq:SmallTauExpFK})), which
prevents us from introducing a local SCET operator for the function
$F-K$. Because this region may be studied by parton shower models
\cite{Sjostrand:1993yb} that respect the exact matrix element
\cite{Catani2001} and because the contribution from the relevant jet
configuration can in principle be removed by double-tag experiments, we
concentrate our attention on the study of the correction factor
$C^{\left( G\right) }$ in this report.

\FIGURE[t]{\epsfig{file=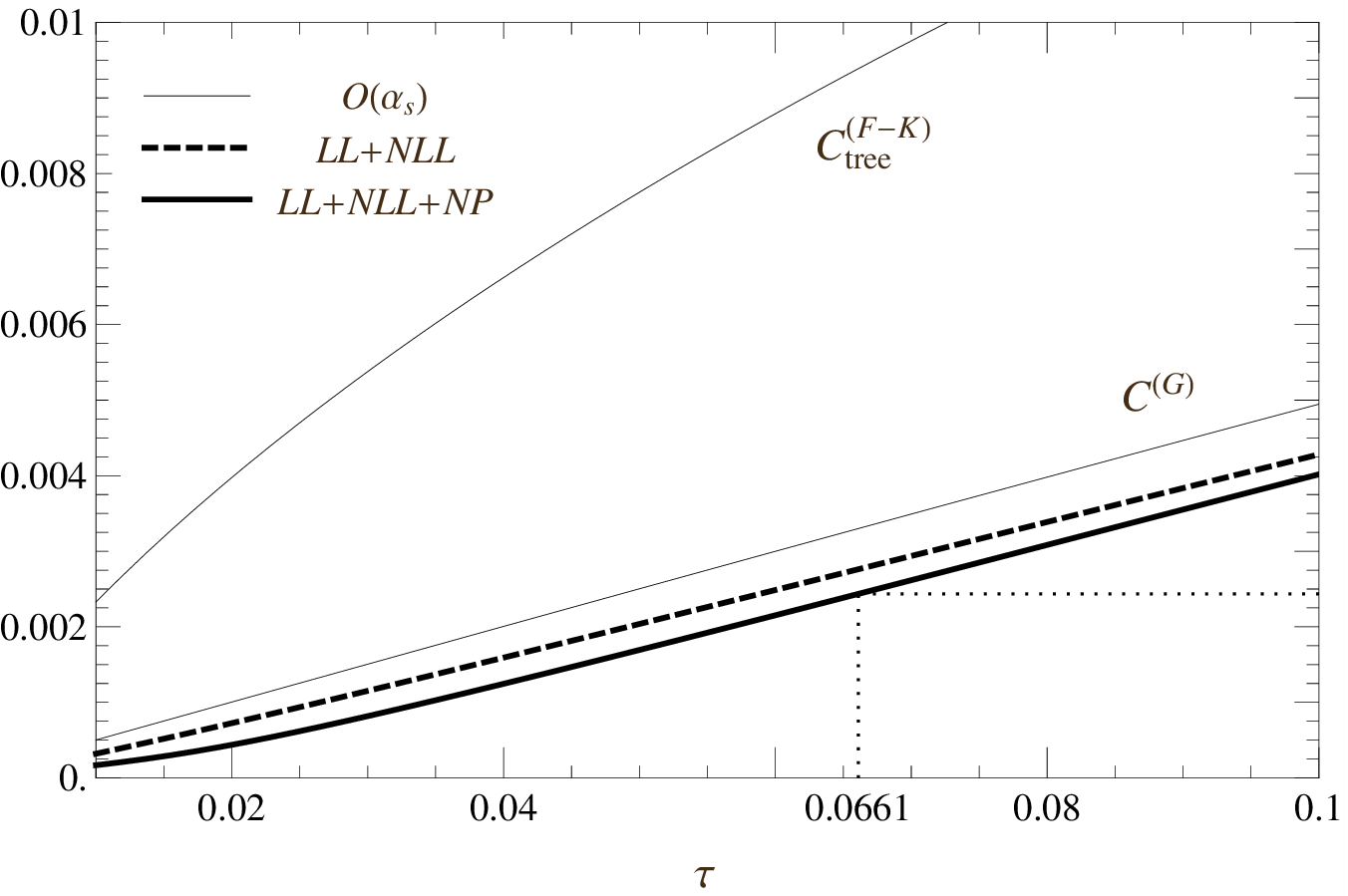,width=.7\textwidth}
\caption{Correction factors for the forward-backward
asymmetry.\label{fig:CCoef}}}

In contrast to $F-K$, the function $G\left(  \tau\right)  $ originates
from the local three-body operator $\Hat{\mathcal{O}}_{3}$ of
Eq.(\ref{eq:O3LC}) in SCET, and we could show in the previous section
that the leading soft singularities from this operator can be resummed
to give the same Sudakov form factor as for the total distribution in
the two-jet limit ($\tau =1-T\rightarrow0$). Because this common
Sudakov factor cancels in the ratio, the thin solid curve for
$C^{\left( G\right)  }$ in \Fig{\ref{fig:CCoef}} may be regarded as the
zeroth-order prediction of perturbative QCD with the leading log (LL)
thrust distribution.

It is worth noting here that this interpretation does not hold for the
$O\left(  \alpha_{s}\right)  $ curve for $C^{\left(  F-K\right)  }$ in
\Fig{\ref{fig:CCoef}}, because the leading Sudakov factor differs from
the $q\bar{q}$ events even in the $\tau\rightarrow0$ limit.

In \Fig{\ref{fig:CCoef}}, our prediction for $C^{\left(  G\right)  }$
in the NLL level is shown by the dashed curve and thick solid curve is
obtained after incorporating the non-perturbative effects by using the
shape function (\ref{eq:uK}). Since the NLL level resummation
supplemented by the non-perturbative correction (\ref{eq:uK})
reproduces the observed thrust distribution rather well (see
Refs.\,\cite{Korchemsky1998,Korchemsky1999} and
\Fig{\ref{fig:DiffThrustDistr}} below), one can directly compare the FB
asymmetry of the observed jets in the two-jet region $1-T< \tau\ll 1$
with the QCD prediction of Eq.\,(\ref{eq:C}) with the correction factor
$C^{\left( G\right)  }\left(  \tau\right)  $ depicted by the thick
solid curve in \Fig{\ref{fig:CCoef}}, provided that the contribution
from \thinspace$F-K$ is reliably excluded by double-tagging and that
the $\tau$-dependance of the observed asymmetry is consistent with the
prediction. If one uses all the events with $T>\left\langle
T\right\rangle = 0.934$ at $Q=m_{Z}$ \cite{Heister:2003aj}, the QCD
correction factor is estimated as
\begin{equation}
\left.  C^{\left(  G\right)  }\left(  \tau\right)  \right\vert _{\tau
= 1-\left\langle T\right\rangle =0.066}=0.0024\pm 0.0002,\label{eq:CMean}
\end{equation}
in the massless quark limit, when the error is estimated by a quadratic
sum of the uncertainty in $\alpha_{s}\left(M_{Z}\right) =0.1184\pm
0.0007$ \cite{Amsler:2008zzb} and the variation due to the different
ans\"atze for the non-perturbative shape function found in
Ref.\thinspace\cite{Korchemsky1998} and
Ref.\thinspace\cite{Korchemsky2000}.

In actual experimental analysis, one needs to correct for effects due
to hadronization, detector acceptance and performances. They have been
accounted for by making use of parton shower based event simulator,
such as JETSET \cite{Sjostrand:1993yb}, which has been tuned to
reproduce all the $e^{+}e^{-}\rightarrow hadrons$ data at collision
c.\thinspace m. energies between 30 and 200~GeV. The difficulty of the
experimental analysis summarized in Ref.\thinspace\cite{Abbaneo1998}
for estimating the electroweak asymmetry parameter $A^{\left(  0\right)
}$ from the observed forward-backward asymmetry of charm and bottom
quark jets may be traced back to the absence of the parton shower that
describes the longitudinal structure function in
Eq.\thinspace(\ref{eq:Gdef}).

It is worth noting here that the traditional scheme \cite{Catani2001}
to match the parton shower (that incorporates resummation of LL and NLL
emissions) and the matrix elements (that give quantum mechanical
correlations such as angular distributions) cannot simulate jets with
correct angular distribution at the accuracy level of precision EW
measurements. This is because in the leading order, the two-jet like
events are matched to the two-parton matrix element which in
$e^{+}e^{-}\rightarrow q\bar{q}$ contributes only to the transverse
structure functions with $1+\cos^{2}\theta_{T}$ and $\cos\theta_{T}$
angular distributions. It is only for the three or more jet events the
tree-level matrix elements contribute to the longitudinal structure
function. Although the longitudinal contributions are power suppressed
in the two-jet region, the accuracy required by the EW precision
measurements may not allow us to neglect their contribution. In this
work we have shown that they can be described as parton shower
originated from the local three-body operator in SCET and that the
resulting jet structure is similar but different from that of the jets
from the $q\bar{q}$ operator. It is a usual consequence of color
coherence that two collinear partons are indistinguishable for a large
angle radiation and should be replaced by a single pseudo-particle in a
jet clustering algorithm. The unusual point specific for power
corrections is that the corresponding three-parton matrix element
should be supplied by only two $\Delta_{q}$ Sudakov form factors and
then combined with two jet-like parton showers. The short distance
nature of longitudinal events requires modify the method to combine
parton showers with matrix elements. A successful Monte-Carlo
simulation of the QCD corrections to the FB asymmetry due to
longitudinal structure function can be carried out only after such
modification is performed.

Before closing this section let us discuss our predictions for the
angular distribution of jets with a particular thrust value. The
differential jet angular distribution can be expressed as
\begin{multline}
\frac{\mathrm{d}\sigma}{\mathrm{d}\cos\theta_{T}\mathrm{d}T}=\frac{3}{4}\,\sigma^{\left(
0\right)}\left[  F^{\prime}\left(  1-T\right) \,\frac{1+\cos
^{2}\theta_{T}}{2}\right.\\ \left.+\frac{4}{3}A^{\left( 0\right)
}K^{\prime}\left( 1-T\right)  \cos\theta_{T}+G^{\prime}\left(
1-T\right)  \sin^{2}\theta _{T}\right]
\end{multline}
in terms of derivatives of the resummed functions at $\tau=1-T$.
Integration
over $\cos\theta_{T}$ gives the thrust distribution:%
\begin{equation}
\frac{\mathrm{d}\sigma}{\mathrm{d}T}=\sigma^{\left(  0\right)  }\left[
F^{\prime}\left(  1-T\right)  +G^{\prime}\left(  1-T\right)  \right]\label{eq:DiffThrust}
\end{equation}
and the FB asymmetry of the jet with a thrust value of $T$ is expressed as%
\begin{align}
\tilde{A}_{FB}\left(  T\right)
&=\frac{\int_{0}^{1}\mathrm{d}\cos\theta_{T}\,\frac{\mathrm{d}\sigma}{\mathrm{d}\cos\theta_{T}\mathrm{d}T}
-\int_{-1}^{0}\mathrm{d}\cos\theta_{T}\,\frac{\mathrm{d}\sigma}{\mathrm{d}\cos\theta_{T}\mathrm{d}T}}
{\int_{-1}^{1}\mathrm{d}\cos\theta_{T}\,\frac{\mathrm{d}\sigma}{\mathrm{d}\cos\theta_{T}\mathrm{d}T}}\notag \\
&=A^{(0)}\frac{K'\left(1-T\right)}{F'\left( 1-T\right)  +G'\left(1-T\right)}.
\end{align}
The correction factor can also be defined for a given $T$ value%
\begin{equation}
\tilde{A}_{FB}\left(  T\right)  =A^{\left(  0\right)  }\left[  1-\tilde
{C}\left(  T\right)  \right]
\end{equation}
with%
\begin{align}
\tilde{C}\left(  T\right)   &  =\tilde{C}^{\left(  F-K\right)  }\left(
T\right)  +\tilde{C}^{\left(  G\right)  }\left(  T\right)  ,\\
\tilde{C}^{\left(  G\right)  }\left(  T\right)   &  =\frac{F^{\prime}\left(
1-T\right)  -K^{\prime}\left(  1-T\right)  }{F^{\prime}\left(  1-T\right)
+G^{\prime}\left(  1-T\right)  },\\
\tilde{C}^{\left(  G\right)  }\left(  T\right)   &  =\frac{G^{\prime}\left(
1-T\right)  }{F^{\prime}\left(  1-T\right)  +G^{\prime}\left(  1-T\right)  }.
\end{align}

As in the case for the asymmetry of jets with $T>1-\tau$, in the
two-jet region where the primary quark and antiquark have momenta in
the opposite hemisphere (see Figs.\,\ref{fig:topo}(a) and (b)), we can
safely assume that $F\left(  \tau\right)  =K\left(  \tau\right)  $ and
all the functions receive common non-perturbative corrections.

The thrust distribution (\ref{eq:DiffThrust}) is shown as the dashed
curve in \Fig{\ref{fig:DiffCorr}} from the resummed expressions
(\ref{eq:R3resum}) and (\ref{eq:R2resum}). The inclusion of the
non-perturbative soft factor (\ref{eq:uK}) shift the prediction to the
solid curve. The LL+NLL prediction supplemented by the non-perturbative
soft factor of Eq.\thinspace(\ref{eq:uK}) reproduces the observed
thrust distribution in \thinspace$e^{+}e^{-}$ collisions at all
energies between 14 and 206~GeV \cite{Stenzel:1999it, Korchemsky1998},
and in particular at $\sqrt{s}=M_{Z}$ gives the mean value $\langle
T\rangle=0.934$ and the distribution is peaked at
$T_{\mathrm{peak}}=0.9794$. Although our evaluation
(\ref{eq:GPertResult}) of $G^{\prime}\left( \tau\right)$ can be
regarded as a part of the NNLO correction to the thrust distribution,
we don't observe significant change in the total thrust distribution.
In \Fig{\ref{fig:DiffCorr}}, we show our prediction for the QCD
correction to the thrust axis angular asymmetry, $\tilde{C}^{\left(
G\right)  }\left(  T\right)  $, for the jets with a particular thrust
value $T$. Again, the dashed curve shows our LL+NLL order prediction
and the solid curve is obtained after including the non-perturbative
soft factor of Eq.\thinspace(\ref{eq:uK}) for both $F$ and $G$
functions. The correction factor $\tilde{C}\left( T\right)  $ decreases
by the non-perturbative correction because it reduces the longitudinal
distribution $G^{\prime}\left(  T\right)  $ more strongly than
$F^{\prime}\left(  T\right)  $ at $T\approx1$. We find
\begin{align}
\tilde{C}^{\left(  G\right)  }\left(  T\right)   &  = 0.0008\qquad
\text{at}\qquad T=T_{\mathrm{peak}}=0.9794,\label{eq:CtildePeak}\\
\tilde{C}^{\left(  G\right)  }\left(  T\right)   &  = 0.008\qquad
\text{at}\qquad T=\left\langle T\right\rangle =0.934,\label{eq:CtildeMean}
\end{align}
from \Fig{\ref{fig:DiffCorr}}. The correction factor is small but it
can be a significant fraction of the error of the $b$-jet FB asymmetry
which is as small as $1.7\%$ on the Z-boson pole \cite{LEP2006}.

Two remarks on our predictions, Eqs.(\ref{eq:CMean}) and
(\ref{eq:CtildeMean}), are in order here: The first is on the validity
of the non-perturbative corrections that lead to our predictions, and
the second is on the additional correction $C^{\left( F-K\right)  }$
due to events where both quark and antiquark are emitted in the same
hemisphere along to the thrust axis opposite the gluon jet direction,
see \Fig{\ref{fig:topo}}(c).

Our prediction for the asymmetry correction factor $C^{\left(  G\right)
}\left(\tau\right)$ and $\tilde{C}^{\left(  G\right)  }\left( T\right)$
are obtained under the assumptions that both the transverse and
longitudinal functions $F\left(  \tau\right)  $ and $G\left(
\tau\right)$, respectively, receive common non-perturbative corrections
via the soft factor Eq.(\ref{eq:uK}), whose form has been chosen to
reproduce the observed jet thrust distribution in
$e^{+}e^{-}\rightarrow hadrons$ experiments. We believe that it is an
excellent approximation in the two-jet ($T\rightarrow1$) limit where
collinear quarks and gluons radiates soft gluons and hadronizes
coherently. Nevertheless, it is clear that the convolution with the
shape function (\ref{eq:uK}) does not exhaust all non-perturbative
effects. Therefore, a careful study by using a shower MC program that
incorporates the longitudinal radiation function $G\left(  \tau\right)
$ is desired for a quantitative estimate.

\DOUBLEFIGURE[t]{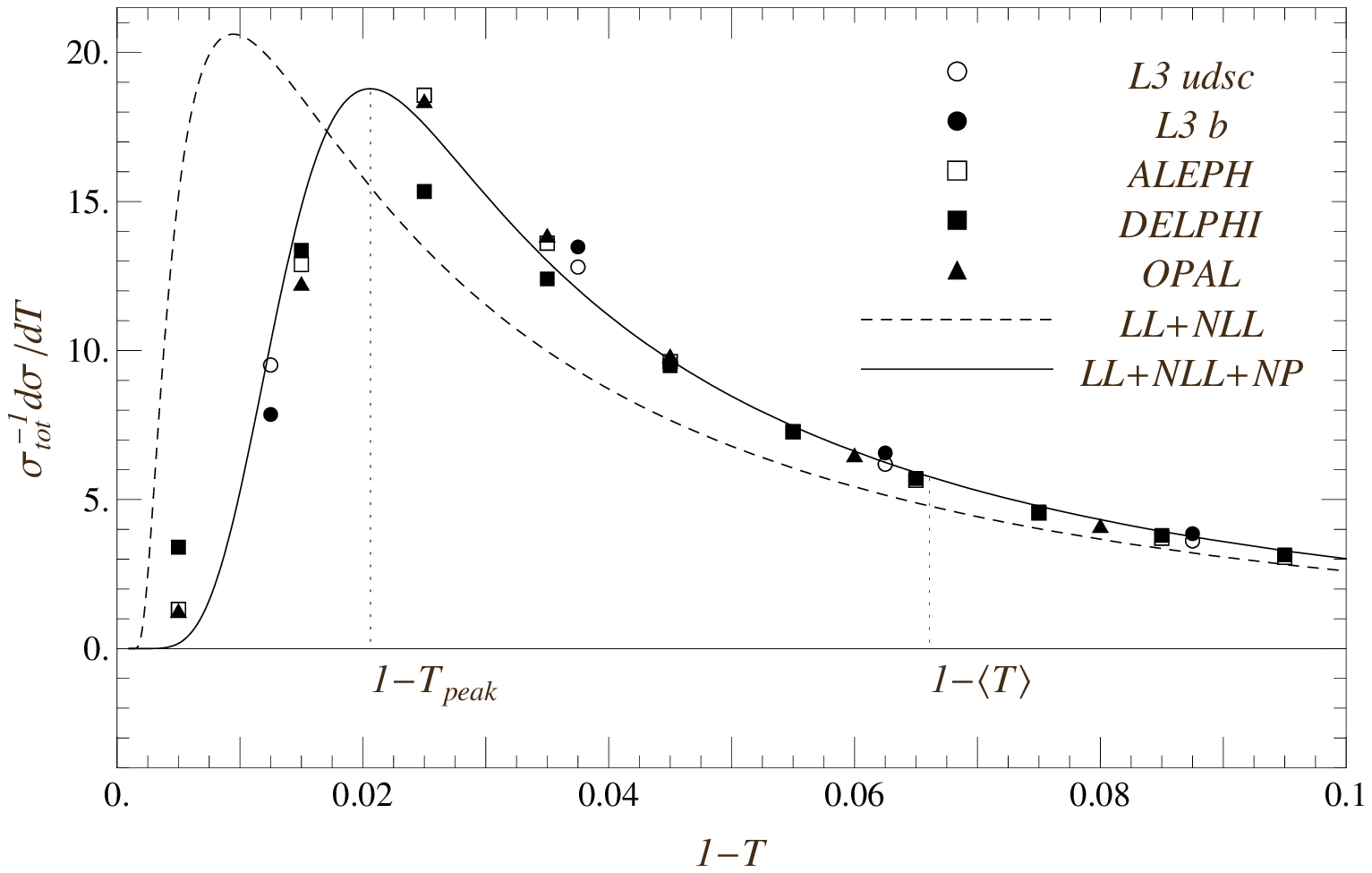,width=.49\textwidth}{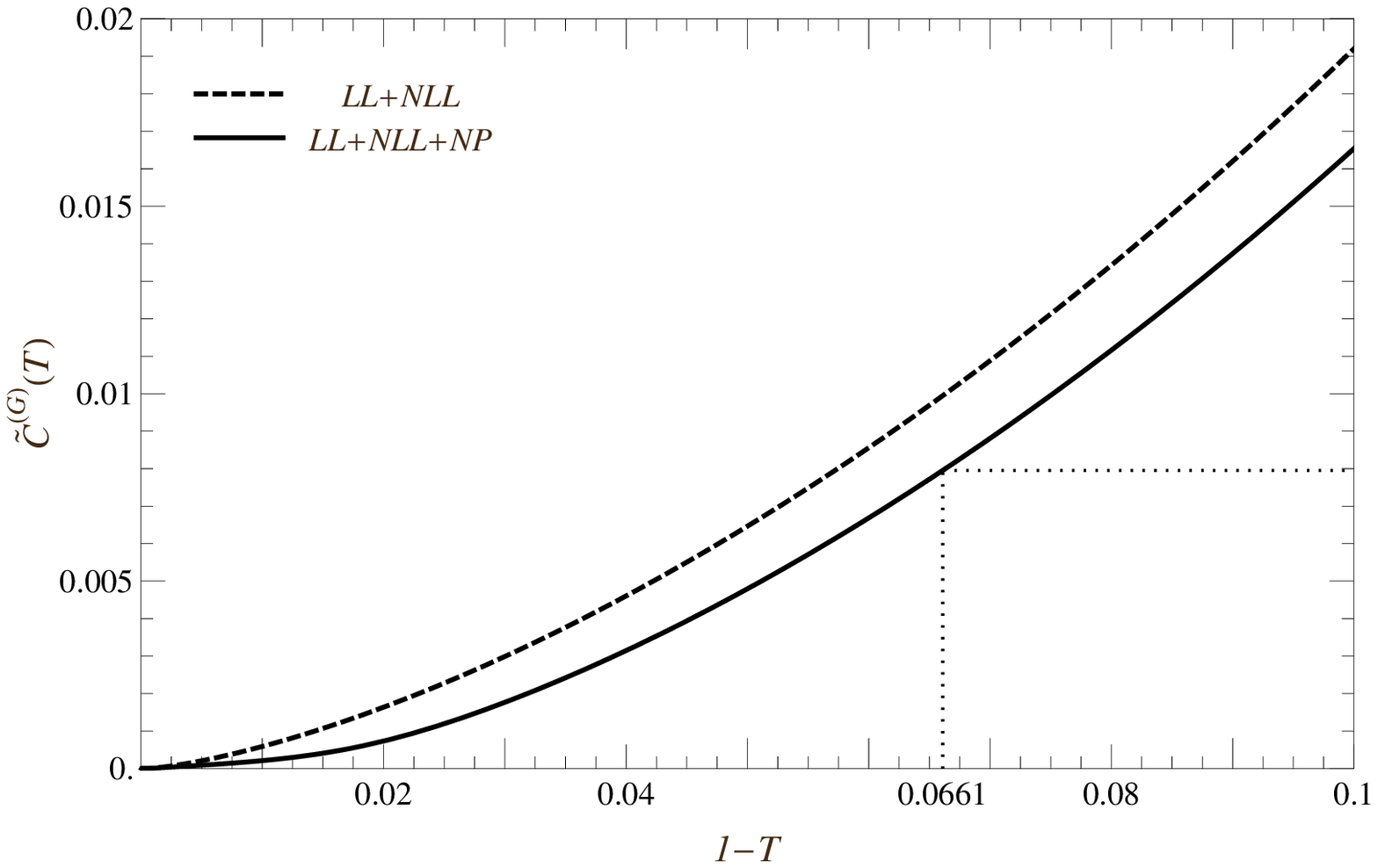,width=.49\textwidth}{Differential
thrust distribution. The data are given in
Refs.\cite{Achard:2004sv,Heister:2003aj,Abreu:2000ck,Abbiendi:2004qz}\label{fig:DiffThrustDistr}}{Corrections
obtained from the differential cross section.\label{fig:DiffCorr}}

In this report we consider mainly the correction due to the
longitudinal function $G\left(\tau\right) $ which arises from the
three-jet configurations Figs.\thinspace\ref{fig:topo}(a),~(b) in the
next-to-leading order, since the contribution to $F-K$ from the
configuration \Fig{\ref{fig:topo}}(c) can be removed by requiring quark
and antiquark momenta are in opposite hemispheres in double flavor-tag
experiments. In practice, however, the double tagging condition leads
to to the reduction of the number of useful events by one order of
magnitude and hence to the loss of accuracy of the measurement. We note
here that probably the present PS program matched to the tree-level
matrix element for three partons \cite{Catani2001} can correctly
account for those events where the quark and the antiquark are in the
same thrust hemisphere. It is easy to show that the matrix element
corresponding to the $F-K$ combination for $1-T\ll1$ is singular in the
$(p_{q}\cdot p_{g})\to 0$ limit, where $p_{q}$, $p_{g}$ are the quark
and gluon momenta, respectively. Due to this singularity, the
logarithmically enhanced contribution comes from the phase-space region
where the energy of the quark is much smaller than that of the
aniquark. For a fixed thrust value such hierarchy leads to a large
opening angle between $q$ and $\bar{q}$. This probably implies that one
can neglect the interference between the radiation emitted along the
quark or the antiquark momentum directions, i.e., three-jet
configurations give a dominant contribution to $F-K$ even for
$1-T\ll1$. However, the additional study of accompanied radiation is
needed. Nevertheless, an extreme care is necessary for a reliable
estimate of the error due to $F-K$, since it is the dominant source of
the correction to the FB asymmetry as shown in \Fig{\ref{fig:CCoef}}.
It mimics the primary quark and antiquark jets in single flavor-tag
experiments. It is not clear to us if all these points have been
appropriately accounted for in the error analysis presented in
Ref.\,\cite{Abbaneo1998}.

\section{Conclusion}

The corrections to the angular distribution of the thrust axis
considered in this paper are power suppressed with respect to the event
shape variable in the two jet region. We identify two mechanisms that
radiation affects the angular distribution in the event topology where
the primary quark and antiquark are radiated into the opposite
hemispheres. It is found that the short-distance process gives the
leading contribution to the longitudinal cross section. Using SCET, we
propose the factorization formula for the longitudinal cross section
and perform the large logarithm resummation to the next-to-leading
logarithmic level of accuracy. The factorization formula allows us to
study the leading nonperturbative corrections to the longitudinal cross
section. A part of the QCD corrections to the forward-backward
asymmetry is the ratio of the longitudinal and total thrust cross
sections. We find that the resummation and the nonperturbative effects
result in additional suppression of this ratio (about $30\%$ for the
experimental mean value of the thrust). We observe that the
short-distance nature of the leading correction yields potential
problems with the Monte-Carlo simulation of the QCD corrections to the
forward-backward asymmetry. We present estimates for the QCD
corrections to the forward-backward asymmetry in the LL+NLL level
including non-perturbative corrections. Underestimation of such
corrections may be relevant for the discrepancy between the weak mixing
parameter $\sin^{2}\theta_{W}$ extracted from the jet asymmetry data
and the others. However, it can be found out only with a help of a new
improved Monte-Carlo simulation.

\section{Acknowledgements}

We thank G.C.~Cho, T.~Gehrmann, T.~Kawamoto for illuminating
discussion. This work is supported in part by Japan Society for
Promotion of Science (JSPS) and in part by Grant-in-Aid for scientific
research (\textnumero 21-09226) from MEXT, Japan. KH wishes to thank
Aspen Center for Physics and the organizers of the 2008 summer workshop
there, where the idea of applying SCET on the $e^{+}e^{-}$ jet angular
distribution was formed.

\bibliographystyle{JHEP}

\end{document}